\documentclass[%
superscriptaddress,
twocolumn,
 amsmath,amssymb,
aps,
prb,
floatfix,
]{revtex4-2}

\usepackage{graphicx}
\usepackage{dcolumn}
\usepackage{booktabs} 
\usepackage{bm}
\usepackage{physics}
\usepackage{amsmath,amsthm,amssymb}
\usepackage{bbold}
\usepackage{hyperref}
\hypersetup{
    colorlinks=true,              
    linkcolor=blue, 
    citecolor=blue,        
    filecolor=magenta,      
    urlcolor=blue           
} 

\usepackage{tikz}
\usetikzlibrary{3d}
\usepackage{slashed}
\usetikzlibrary{calc,decorations.markings}

\newcommand{\vk}{\mathbf{k}}
\newcommand{\vq}{\mathbf{q}}

\newcommand{\ek}{\varepsilon_{\lambda,\vk}}

\renewcommand{\tilde}{\widetilde}              
\renewcommand{\vec}[1]{\mathbf{#1}}            
\newcommand{\sgn}{\textup{sgn}}                

\begin{document}

\title{Phonon assisted absorption in Transition Metal Dichalcogenide heterostructures}


\author{Yifan Liu}
\affiliation{Department of Physics \& Astronomy, University of California, Riverside, CA 92521, USA}

\author{Robert Dawson}
\affiliation{Department of Physics, University of Notre Dame, Notre Dame, IN 46556, USA}

\author{Nathaniel Gabor}
\affiliation{Department of Physics \& Astronomy, University of California, Riverside, CA 92521, USA}

\author{Vivek Aji}
\email{vivek.aji@ucr.edu}
\affiliation{Department of Physics \& Astronomy, University of California, Riverside, CA 92521, USA}

\date{\today}

\begin{abstract}
    The coupling of atomic vibrations to electronic excitations --- traditionally understood to be a source of energy loss in semiconductors --- has recently been explored in photosynthetic light harvesting as a means to circumvent dissipation by harnessing quantum vibronic coherence. Motivated by recent photocurrent measurements of vibronic sidebands in WSe2/MoSe2 optoelectronic devices [F. Barati \textit{et al.}, \textit{Nano Lett.}\ \textbf{22}, 5751 (2022)], we present a nonperturbative theoretical framework for phonon-assisted absorption in van der Waals heterostructures. Using a polaron transformation, a closed-form expression for the optical absorption spectrum at arbitrary temperatures is presented. Our model includes both intraband and interband electron–phonon coupling. Detailed analysis shows that the observed periodic sidebands originate from the strong coupling between interlayer excitons and nearly dispersionless optical phonon modes. By comparing two limiting cases, one involving only intraband couplings and the other incorporating coherent interband processes, we show that interband phonon-assisted transitions are needed to account for the observed data. Beyond enabling the direct estimation of vibronic coupling strengths from spectroscopic data, these findings have profound consequences for our understanding of optical and optoelectronic responses: coherent interband coupling of atomic vibrations to excitons is essential to quantifying photoresponse in transition metal dichalcogenide heterostructures.
\end{abstract}

\maketitle

\section{Introduction}

Ultrafast optoelectronic processes are central to a wide range of phenomena, including photosynthesis, photovoltaics, high-speed classical and quantum computing, and time-resolved spectroscopy. All of these rely on the ability to control and manipulate excitations generated by the interaction of light and matter. Energy transferred to charge degrees of freedom, such as electron-hole pairs and excitons, by photon absorption follows multiple pathways as the system relaxes toward the final state. Navigating this complex landscape, shaped by equilibration of the electron-hole system within and between bands or valleys, dissipation via phonon generation and inelastic impurity scattering, and recombination, is both a challenge and an opportunity for achieving improved functionality. The large number of degrees of freedom, multiple competing time and length scales, and limited direct experimental access to the fundamental dynamical processes involved are some of the key roadblocks to be overcome.

Biomimetics has emerged as an innovative way to explore the design space of ultrafast optoelectronics. In particular, photosynthetic light-harvesting systems are highly efficient in absorbing incident radiation despite the fluctuating and noisy environments in which they operate~\cite{doi:10.1126/science.1142188,10.1063/1.3002335,fei_nat_comm_19}. The coupling of molecular motion to exciton dynamics, which traditionally is viewed as a source of loss, can enhance energy transfer. Translating this idea to crystalline systems suggests that exciton-phonon quasiparticles, the solid-state analog of molecular vibronic states, may provide pathways that circumvent dissipative losses~\cite{toyozawa, PhysRevLett.94.027402, leturcq}. A signature of such excitations in molecular systems is the existence of periodic vibronic peaks in absorption spectra~\cite{harris}. Recent advances in layered semiconducting devices have shown early promise in realizing analogous phenomena in photodiodes~\cite{PhononData}.

Another aspect that has enabled progress in optoelectronics is the development of photocurrent measurements. Traditionally, photoluminescence (PL) and pump-probe techniques have been the basis of much of our understanding of light-matter interaction~\cite{RevModPhys.74.895}. In both cases, the underlying dynamical processes cannot be directly accessed except in special circumstances. As a result, ultrafast phenomena have to be inferred by modeling how the initial excitation transfers energy to other degrees of freedom before recombination, and by fitting to the observed PL line shape or changes in the absorption or reflection of a probe beam. Coherent excitations within the electronic sector have also been studied using techniques such as four-wave mixing and coherent emission detection~\cite{RevModPhys.74.895}, and optical generation of coherent phonons coupled to other excitations such as plasmons~\cite{PhysRevLett.77.4062} and Bloch oscillations~\cite{dekorsy} provides one route to accessing vibronic phenomena. A more generic approach, avoiding the need for coherent phonons, is the direct coupling of excitons to phonons. Photocurrent measurements allow for the detection of phonon-assisted absorption by measuring the photogenerated charges as a function of incident energy.

In this paper, we analyze the vibronic sidebands observed in the photocurrent generated in a layered semiconductor photodiode~\cite{PhononData}. A van der Waals heterostructure consisting of bilayer tungsten diselenide (WSe$_2$) stacked on monolayer molybdenum diselenide (MoSe$_2$) displays multiple periodic and well-separated photocurrent peaks, as a function of incident photon energy, around the interlayer exciton energy. These are hard to detect in PL due to the weak interlayer exciton oscillator strength~\cite{rivera}. Furthermore, the sidebands are not observed in the photocurrent spectra near the intralayer exciton energies. The peaks are separated by 30 meV, which matches the window of a narrow branch of phonons expected for the interlayer heterostructure. Since the electron--hole pairs produced in the two layers flow to the source and drain electrodes, the measured current is proportional to the net absorption. 

To calculate the absorption coefficient, we employ a nonperturbative framework for phonon-assisted absorption in TMD heterostructures. Our model accounts for both intra- and interband electron--phonon interactions and incorporates a polaron transformation~\cite{PhysRev.90.297, PhysRevLett.21.1637,Feldtmann} that enables analytic control of vibronic effects at arbitrary temperatures. We present two limiting cases: Model I, where phonons couple only to intraband transitions, and Model II, which includes interband coupling through a minimal matrix ansatz. In both cases, we derive absorption spectra in closed form and fit them to experimental data. The comparison reveals that including interband coupling improves the agreement with experimental sideband structures and allows for systematic extraction of effective phonon coupling strengths.

\section{The model}
\begin{figure}
  \centering
  \includegraphics[width=0.5\textwidth]{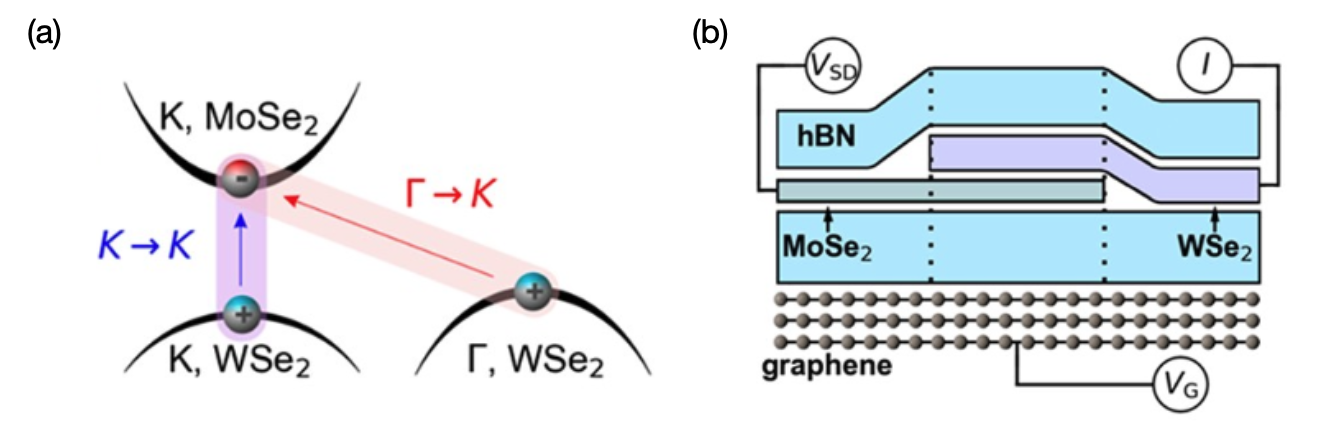}
  \caption{
    (a) Schematic of interlayer excitons formed via $K\rightarrow K$ and $\Gamma\rightarrow K$ transitions in WSe$_2$/MoSe$_2$ heterostructure.  
    (b) Device layout showing layered structure, graphene gate, and contacts. Reproduced with permission from Ref.~[\onlinecite{PhononData}].
  } 
  \label{fig:fig1}
\end{figure}

\begin{figure}
  \centering
  \includegraphics[width=0.48\textwidth]{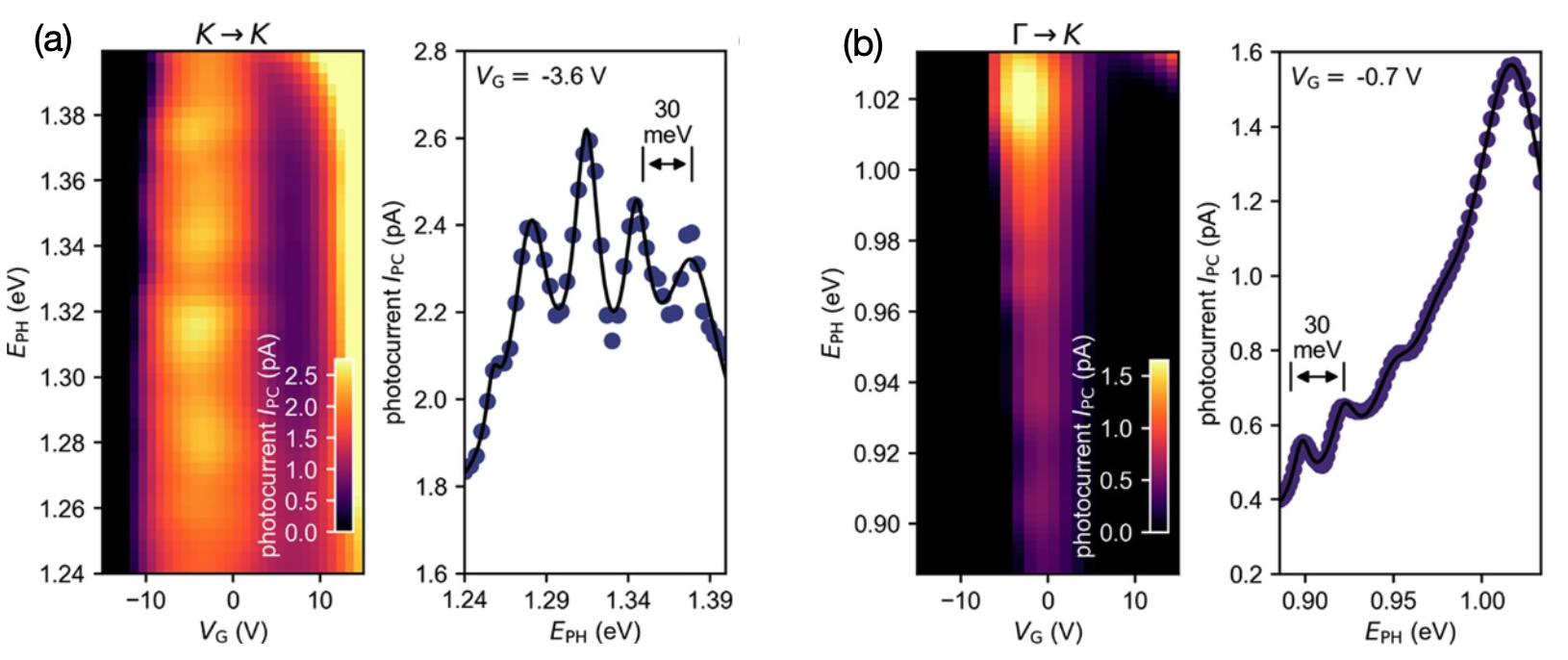}
  \caption{
    Photocurrent spectra revealing periodic vibronic sidebands.  
    (a) Spectrum associated with $K \rightarrow K$ interlayer excitons, showing a series of regularly spaced absorption peaks separated by approximately 30 meV.  
    (b) Gate-dependent photocurrent $I_{\text{PC}}$ near the $\Gamma \rightarrow K$ interlayer exciton resonance, also exhibiting periodic sidebands with similar spacing. Reproduced with permission from Ref.~[\onlinecite{PhononData}].
  }\label{fig:fig2}
\end{figure}

The system under study consists of bilayer WSe$_2$ stacked on monolayer MoSe$_2$, encapsulated between hexagonal boron nitride (hBN) layers and contacted by metal source–drain electrodes and a multilayer graphene gate~\cite{PhononData} (see Fig.~\ref{fig:fig1}b). This vertically aligned stacking supports the formation of tightly bound interlayer excitons, wherein electrons reside in the MoSe$_2$ conduction band at the $K$ point, while holes occupy the WSe$_2$ valence band at either the $K$ or $\Gamma$ point (Fig.~\ref{fig:fig1}a). Owing to the spatial separation between electron and hole, interlayer excitons possess a large intrinsic dipole moment oriented perpendicular to the layers, rendering them sensitive to applied out-of-plane electric fields.
When illuminated by near-infrared light, these excitons can be dissociated by a gate-tunable source–drain bias $V_{\text{SD}}$, giving rise to a photocurrent that serves as a spectroscopic probe of their absorption properties. Remarkably, photocurrent spectra reveal a ladder of sidebands near the $K \rightarrow K$ interlayer exciton resonance, with an energy spacing of approximately 30 meV (Fig.~\ref{fig:fig2}a). A similar, and less prominent, structure is also observed near the $\Gamma \rightarrow K$ interlayer exciton transition (Fig.~\ref{fig:fig2}b). 

\begin{figure}
  \centering
  \includegraphics[width=1\columnwidth]{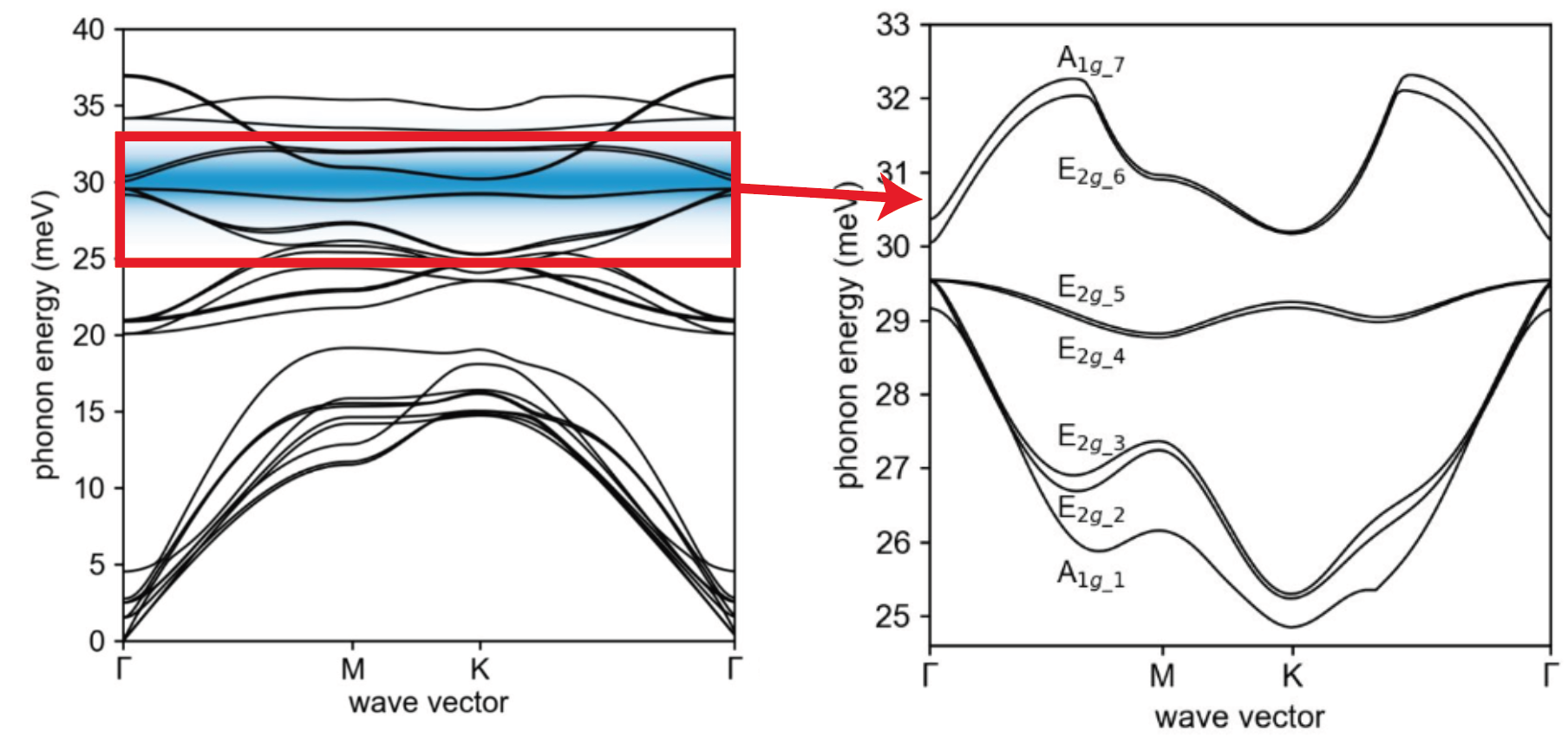}
  \caption{
    Phonon band structure of the heterostructure. The boxed region highlights nearly dispersionless optical phonon modes around 30 meV, which mediate electron–phonon coupling. Reproduced with permission from Ref.~[\onlinecite{PhononData}].
  }
  \label{fig:fig3}
\end{figure}

These sidebands are absent in photoluminescence spectra and do not appear for intralayer excitons in the photocurrent spectra~\cite{PhononData}. Their uniform 30~meV spacing and comparable intensities across multiple phonon quanta indicate a strong, non-perturbative coupling between interlayer excitons and a specific subset of phonon modes with appropriate symmetry and momentum. First-principles calculations~\cite{PhononData} identify a cluster of nearly dispersionless optical phonons around 30~meV, involving both interlayer shear and layer breathing modes, as the most likely candidates for mediating this coupling (Fig.~\ref{fig:fig3}). These phonon modes enable the emergence of periodic sidebands that remain well-resolved within a finite window of gate-controlled bias, reflecting coherent exciton–phonon coupling.

\begin{figure}
\centering
\includegraphics[width=\linewidth]{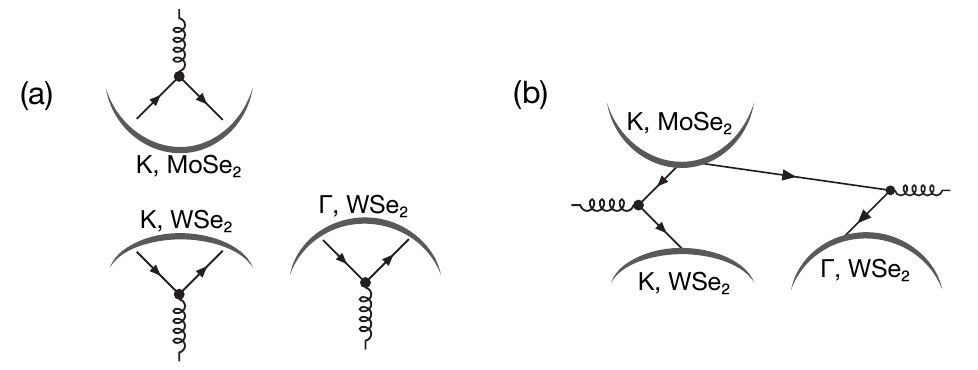}
\caption{
Schematic interaction vertices for electron–phonon coupling processes relevant to our models.  
(a) Model I includes only intraband phonon-assisted processes, where electrons scatter within the conduction or valence band at fixed valley while emitting or absorbing a phonon.  
(b) Model II additionally incorporates interband phonon-assisted transitions between the conduction and valence bands near either the $K$ or $\Gamma$ point. The Hermitian conjugate process is implied but not shown.
}
\label{fig:intra_inter_scattering}
\end{figure}

Such observations raise fundamental questions about the nature of exciton–phonon coupling in TMD heterostructures, and challenge conventional weak-coupling approaches that treat phonons as perturbative broadening mechanisms. To address these questions, we develop in the following sections a microscopic theory of phonon-assisted absorption based on a nonperturbative polaron formalism. This framework allows us to capture the formation of vibronic exciton–phonon states and the emergence of ladder-like absorption peaks, in quantitative agreement with experiment.

To distinguish the microscopic content of the two models studied, Fig.~\ref{fig:intra_inter_scattering} illustrates the allowed electron–phonon interaction processes. Model I includes only intraband phonon-assisted transitions—namely, scattering events where an electron remains within the same band (conduction or valence) and same valley while coupling to a phonon, independently for the $K\rightarrow K$ and $\Gamma\rightarrow K$ exciton processes. Model II, in contrast, incorporates additional interband phonon-assisted transitions between the conduction and valence bands, which allow phonon-mediated processes that change the electron’s band index during the transition. These interband processes are absent in Model I and are important to capturing the quantitative behavior of the absorption features observed in experiment.

\subsection{Model I: Phonon-assisted intraband absorption}
The interlayer exciton is composed of an electron at the $K$ point of the MoSe$_2$ conduction band and a hole at either the $K$ or $\Gamma$ point of the WSe$_2$ valence band. In addition to the electron-hole Coulomb attraction, the charges are coupled to phonons. In particular there exists a collection of 7 interlayer lattice vibrational modes with narrow bandwidth centered at $\sim$30~meV~\cite{PhononData}. The antisymmetric modes allow for coupling to the conduction and valence bands to have opposite sign. The small bandwidth implies that the dispersion of the phonons is negligible across the Brillouin zone as compared to the electronic degrees of freedom and that the modes are spatially localized. A general model that  captures the effects of electron-electron, phonon-electron, and photon-electron interactions is:
\begin{align}
    H_0 &= \sum_{\lambda,\vk}\ek P^{\lambda,\lambda}_{\vk,\vk} + \sum_{\nu,\vq}\hbar\Omega^\nu_\vq(D^\dagger_{\nu,\vq}D_{\nu,\vq}+\tfrac{1}{2})\nonumber\\
    &\,+ \sum_{\vq}\hbar\omega_\vq(B^\dagger_\vq B_\vq + \tfrac{1}{2})\\
    H_{el-el} &= \frac{1}{2}\sum_{\lambda,\lambda'}\sum_{\vk,\vk'}\sum_{\vq\neq 0} V^{\lambda,\lambda'}_\vq a^\dagger_{\lambda,\vk}a^{\dagger}_{\lambda',\vk'}a_{\lambda',\vk'+\vq}a_{\lambda,\vk-\vq}\\
    H_{el-ph} &= \sum_{\lambda,\vk}\sum_{\nu,\vq}\hbar\Omega^{\nu}_\vq g^{\lambda}_{\nu,\vq} P^{\lambda,\lambda}_{\vk-\vq,\vk}(D^\dagger_{\nu,\vq} + D_{\nu,-\vq})\\
    H_{el-em} &= -i\sum_{\lambda,\lambda'}\sum_{\vk}\mathcal{F}^{\lambda,\lambda'}P^{\lambda,\lambda'}_{\vk,\vk}B^{\lambda,\lambda'}_\vk,
\end{align}
where $a^\dagger_{\lambda,\vk} (a_{\lambda,\vk})$ creates (annihilates) an electron with momentum $\vk$ and energy $\ek$ in band $\lambda$, $D^\dagger_{\nu,\vq} (D_{\nu,\vq})$ creates (annihilates) a phonon with momentum $\vq$ and frequency $\Omega^{\nu}_\vq$ in branch $\nu$, and $B^\dagger_{\vq} (B_{\vq})$ creates (annihilates) a photon with momentum $\vq$ and frequency $\omega_\vq$. Additionally, we have introduced shorthand notation for the electronic polarization operator $P^{\lambda,\lambda'}_{\vk,\vk'}\equiv a^\dagger_{\lambda,\vk}a_{\lambda',\vk'}$, and the operator $B^{\lambda,\lambda'}_\vk$ is defined to be $B_{\vq=0}$ for $\varepsilon_{\lambda,\vk}>\varepsilon_{\lambda',\vk}$ and $B^\dagger_{\vq=0}$ for $\varepsilon_{\lambda,\vk}<\varepsilon_{\lambda',\vk}$. The matrix elements $g^{\lambda}_{\nu,\vq}$ and $\mathcal{F}^{\lambda,\lambda'}$ characterize the Fr\"ohlich and electromagnetic interactions, respectively. 

Equations of motion for the polarization and field theoretic evaluation of the retarded Green’s function are two approaches that are typically employed to calculate the optical absorption. A brief overview and the main results are presented here. Technical derivations are provided in Appendices~\ref{ap_eom} and~\ref{ap_gf}.

The photon absorption/emission spectra can be obtained through the equation of motion of the electron polarization operator $P^{\lambda,\lambda'}_\vk$~\cite{doi:10.1142/7184}. However, to obtain the phonon side-bands in the spectra, the equations result in an infinite hierarchy. In Ref.~[\onlinecite{Feldtmann}] it was shown that the interacting polaron picture is a more suitable framework. Generalizing the approach we include the phonon branch into the unitary transformation: $U=\exp\bigg[\sum_{\nu,\vq}\sum_{\lambda,\vk}g^{\lambda}_{\nu,\vq} P^{\lambda,\lambda}_{\vk-\vq,\vk}Q^\nu_\vq\bigg]$
where $Q^\nu_\vq\equiv D^\dagger_{\nu,\vq}-D_{\nu,-\vq}$. Applying the Baker–Campbell–
Hausdorff formula,  we obtain the interacting polaron picture:
\begin{align}
    \label{eq:Pol}
    \bar{H}_{pol} &= \sum_{\lambda,\vk,\vq}\left[\bigg(e^{-\mathcal C_\lambda}\mathcal{E}_\lambda e^{\mathcal C_\lambda}\bigg)_{\vk,\vk-\vq}P^{\lambda,\lambda}_{\vk-\vq,\vk} 
    \right. \nonumber\\
    &\qquad-\left.\bigg(\sum_{\nu}\hbar\Omega^\nu_{\vq}|g^{\lambda}_{\nu,\vq}|^2\bigg)P^{\lambda,\lambda}_{\vk,\vk}\right]\\
    \bar{H}_{latt} &= \sum_{\nu,\vq}\hbar\Omega^\nu_\vq(D^\dagger_{\nu,\vq}D_{\nu,\vq}+\tfrac{1}{2})\\
    \bar{H}_{em} &= \sum_{\vq}\hbar\omega_\vq(B^\dagger_\vq B_\vq + \tfrac{1}{2})\\
    \bar{H}_{pol-pol} &= \tfrac{1}{2}\sum_{\lambda,\lambda'}\sum_{\vk,\vk'}\sum_\vq \tilde{V}^{\lambda,\lambda'}_\vq a^\dagger_{\lambda,\vk}a^\dagger_{\lambda',\vk'}a_{\lambda',\vk'+\vq}a_{\lambda,\vk-\vq}\\
    \label{eq:Pol-em}
    \bar{H}_{pol-em} &= -i\sum_{\lambda,\lambda'}\mathcal{F}^{\lambda,\lambda'}\bigg(e^{\mathcal C_\lambda-\mathcal C_{\lambda'}}\bigg)_{\vq,0}P^{\lambda,\lambda'}_{\vq,0}B^{\lambda,\lambda'}
\end{align}
where we have adopted the same compact matrix notation as in [\onlinecite{Feldtmann}], such that the momentum component of a product of two matrices $M$ and $N$ is given by $(MN)_{\vk,\vk'}=\sum_{\vk_1}M_{\vk,\vk_1}N_{\vk_1,\vk'}$. The phonon operator dynamics are contained within the operator $\mathcal C^\lambda_{\vk-\vk'}\equiv\sum_\nu g^{\lambda}_{\nu,\vk-\vk'}Q^\nu_{\vk-\vk'}$, and $\tilde{V}^{\lambda,\lambda'}_\vq$ is a modified Coulomb interaction given by $\tilde{V}^{\lambda,\lambda'}_\vq = V^{\lambda,\lambda'}_\vq - \sum_\nu 2\hbar\Omega^\nu_\vq (g^{\lambda}_{\nu,\vq})^*g^{\lambda'}_{\nu,\vq}$.
Replacing the phonon operators with their thermal average in Eq.~(\ref{eq:Pol}) leads to an effective electronic band structure $\bar{H}_{pol} =\sum_{\lambda,\vk}e_{\lambda,\vk}P^{\lambda,\lambda}_{\vk,\vk}$, where the explicit form of the polaron dispersion $e_{\lambda,\vk}$ is given in Appendix~\ref{ap_mod1} and its specific form is not essential for the analysis that follows.
The equation of motion for the photon occupation number reads
\begin{align}
\expval{\partial_t\!\left(B_{\vq}^\dagger B_{\vq}\right)}
= -\frac{i}{\hbar}\Big(A_{\vq}(t)-L_{\vq}(t)\Big),
\end{align}
where $A_{\vq}(t)$ and $L_{\vq}(t)$ describe phonon-assisted absorption and emission processes, respectively. Explicitly,
\begin{align}
A_{\vq}(t) &=
\sum_{\varepsilon_\lambda > \varepsilon_{\lambda'}}\sum_{\vk,\vq'}
\mathcal{F}^{\lambda,\lambda'}
\big(e^{\mathcal C_\lambda-\mathcal C_{\lambda'}}\big)_{\vq',0}
P^{\lambda,\lambda'}_{\vq',0}
B_{\vq}\,\delta_{\vq,0},
\label{eq:Absorption}
\\
L_{\vq}(t) &=
\sum_{\varepsilon_\lambda < \varepsilon_{\lambda'}}\sum_{\vk,\vq'}
\mathcal{F}^{\lambda,\lambda'}
\big(e^{\mathcal C_\lambda-\mathcal C_{\lambda'}}\big)_{\vq',0}
P^{\lambda,\lambda'}_{\vq',0}
B^\dagger_{\vq}\,\delta_{\vq,0}.
\label{eq:Emission}
\end{align}
The absorption coefficient is then defined as
\begin{align}
\alpha(\omega) \equiv
\Re\!\left[A(\omega)/\expval{B^\dagger B}\right],
\end{align}
with the explicit expression derived in Appendix~\ref{ap_eom}.

One can also calculate the absorption coefficient at finite temperatures through the greater part of the retarded four-point correlation function of electron operators. Within the framework of linear response theory the carrier-photon coupling terms acts as an external perturbation. The time-evolution of electronic operators is governed by the Hamiltonian without $H_{el-em}$ and $H^0_{em}$. To calculate the correlation functions in an analytically tractable manner we leverage the polaron transformation to treat the electron-phonon coupling nonperturbatively ~\cite{hannewald2005nonperturbative,marsusi2012theoretical}. This approach allows us to factorize the electron and phonon operators and perform thermal averaging on them independently.  From now on, $H$ denotes the full Hamiltonian without $H_{e-em}$ and $H^0_{em}$, i.e. $H=H^0_e+H^0_{ph}+H_{e-e}+H_{e-ph}$, unless otherwise stated. The four point function is (see Appendix~\ref{ap_gf})
\begin{align}
  P^R_>(t)
  &=-i\frac{\Theta(t)}{N}\sum_{\vec {k_1...k_4}}
  \langle
  a^\dagger_{\vec k_1 v}(t) a_{\vec k_2 c}(t) a^\dagger_{\vec k_3 c}(0) a_{\vec k_4 v} (0)
  \rangle \nonumber\\
  &\quad\times \langle 
  (e^{ \mathcal C^v(t)- \mathcal C^c(t)})_{\vec {k_2k_1}} (e^{ \mathcal C^c(0)-  \mathcal C^v(0)})_{\vec {k_4k_3}}
  \rangle,
\end{align}
where polaron operator evolves according to $a^{(\dagger)}_{\vec k \lambda}(t)=e^{i\bar H_{p}t} a^{(\dagger)}_{\vec k \lambda}e^{-i\bar H_{p}t}$ and $\bar H_p=\bar H_{pol}+\bar H_{pol-pol}$, and similarly $ \mathcal C^\lambda(t)=e^{i\bar H_{ph}t} \mathcal  C^\lambda(0)e^{-i\bar H_{ph}t}$ with matrix elements being $   \mathcal C^\lambda(t)_{\vec{kk'}}=\sum_\alpha g^\lambda_{\vec{k-k'},\alpha}(e^{i\Omega_\alpha t}D^\dagger_{\vec{k-k'},\alpha}-e^{-i\Omega_\alpha t}D_{\vec{k'-k},\alpha})$; $N$ is the number of unit cells. 

Eliminating the electron-phonon coupling through the canonical transformation makes it harder to evaluate the broadening of the absorption spectral lines induced by the residual polaron-phonon interaction~\cite{marsusi2012theoretical}. Therefore we treat the broadening as a parameter to be determined by fitting experimental measurements. The absorption coefficient is given by
\begin{align}
  \alpha(\omega,T)
  =-\frac{1}{\pi}\Im P^R_>(\omega,T).
  \end{align}
Evaluating at finite temperatures assuming a single phonon branch, with flat dispersion $\Omega_{\vec k}=\Omega$, and constant coupling constants we obtain (see Appendix~\ref{ap_gf})
\begin{align}
  \alpha(\omega,T)
  &=e^{-g_{cv}^2(2n_B(\Omega,T)+1)}
  \sum_{\vec Q\nu} \sum_{m\in\mathbb{Z}}
  \nonumber\\
  &\frac{e^{\frac{1}{2}m\beta\Omega}|\psi_\nu(\vec 0)|^2\gamma_m}{(\omega-m\Omega-E_{\vec Q\nu})^2+\gamma_m^2}
  I_m\left(\frac{g_{cv}^2}{\sinh(\beta\Omega/2)}\right).\label{eq:Absorption_T}
\end{align}
The various quantities appearing in Eq.(\ref{eq:Absorption_T}) are defined as follows. The coupling constant $g_{cv} = g^c - g^v$ represents the difference between the conduction and valence band couplings to phonons. The Bose-Einstein distribution function is given by $n_B(\Omega,T) = (e^{\beta\Omega} - 1)^{-1}$. Here, $\psi_\nu(\vec{r})$ is the relative-motion wavefunction of the polaron-anti-polaron pair with dispersion $E_{\vec{Q}\nu}$, where $\vec Q$ is the center-of-mass momentum and $\nu$ is a discrete quantum number . $\gamma_m$ is an $m$-dependent aforementioned broadening parameter. Finally, $I_m(z)$ represents the modified Bessel function of the first kind.

\subsection{Model II: Phonon-assisted intraband and interband absorption}
We now extend our analysis to the case where phonons couple not only to intraband electronic transitions but also to interband transitions. In this scenario, the electron-phonon interaction is described by the Hamiltonian: $H_{e-ph}=\sum_{\vec {kq} \lambda \lambda'\alpha}\Omega_{\vec q\alpha}g^{\lambda\lambda'}_{\vec q\alpha}a^\dagger_{\vec {k-q},\lambda'}a_{\vec {k}\lambda}(D_{-\vec q,\alpha}+D^\dagger_{\vec q\alpha})$. The electron-phonon coupling is no longer diagonal in the band index $\lambda$. The non-perturbative treatment of this interaction leads to a significantly more complex transformed Hamiltonian. The proliferation of additional terms render the calculation of the correlation function intractable within a general framework. For shear and breathing modes where the intraband band couplings have equal magnitude and opposite signs for the two bands, with the interband coupling being equal by symmetry, the model is analytically solvable. For this case the coupling takes the form: $g_\alpha=g_1 \sigma_1+g_3\sigma_3$, where $g_1,g_3\in\mathbb R$, $\sigma_1$ and $\sigma_3$ are Pauli matrices acting in the conduction-valence band space. The transformed Hamiltonian retains a structure analogous to the purely intraband case (see Appendix~\ref{ap_gf2}), allowing us to derive an absorption coefficient with a similar functional form. The presence of the off-diagonal term  $g_1 \sigma_1$  introduces additional phonon-assisted transition channels, modifying the effective coupling strength and leading to spectral weights that differ from those observed in the diagonal case.

\begin{figure}
  \centering
  \includegraphics[width=\linewidth]{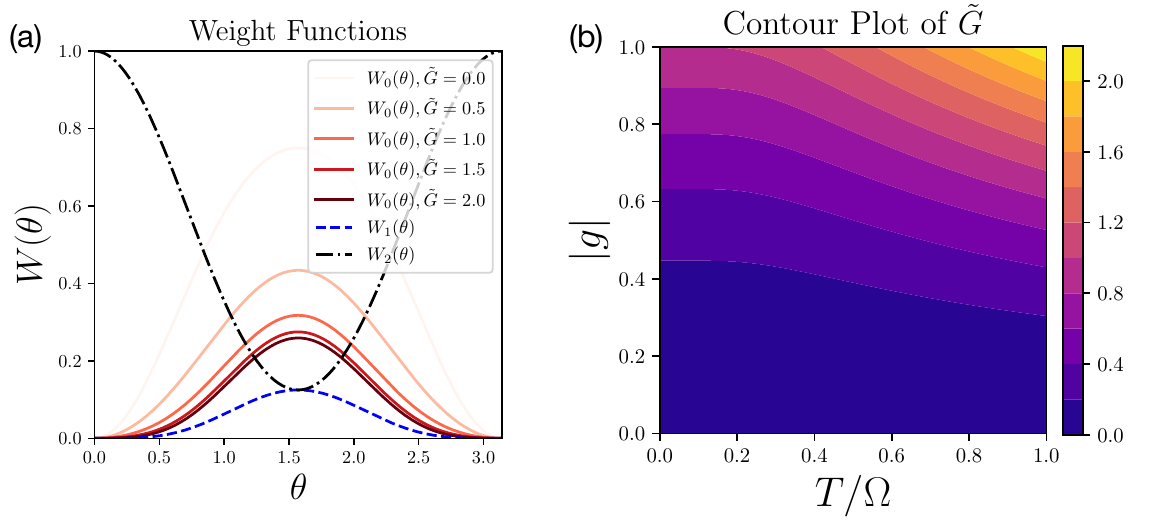}
  \caption{(a) Weight functions $W(\theta)$ for various $\tilde G$ values. (b) Contour plot of $\tilde G$ in the $(T/\Omega,|g|)$ plane. Here $T$ and $\Omega$ are both measured in eV.}
  \label{fig:Weight_functions}
\end{figure}

The transformed greater four-point correlation function in this case reads
\begin{align}
  P^R_>(t)
  &=
  -i\frac{\Theta(t)}{N}
  \sum_{\vec{k},\vec{k}',\{\vec{k}_i\lambda_i\}}
  \langle
  a^\dagger_{\vec k_1 \lambda_1}(t) 
  a_{\vec{k}_2\lambda_2}(t) 
  a^\dagger_{\vec k_3 \lambda_3} 
  a_{\vec k_4 \lambda_4}
  \rangle
  \nonumber\\
  &\times
  \langle 
  (e^{\mathcal C(t)})_{\vec{k}v,\vec{k}_1\lambda_1} 
  (e^{-\mathcal C(t)})_{\vec{k}_2\lambda_2,\vec{k}c}
  \nonumber\\
  &\qquad\qquad
  (e^{\mathcal C(0)})_{\vec{k}'c,\vec{k}_3\lambda_3} 
  (e^{-\mathcal C(0)})_{\vec{k}_4\lambda_4,\vec{k}'v} 
  \rangle
\end{align}
where polaron operator evolves according to $a^{(\dagger)}_{\vec k \lambda}(t)=e^{i\bar H_{p}t} a^{(\dagger)}_{\vec k \lambda}e^{-i\bar H_{p}t}$ and $\bar H_p=\bar H_{pol}+\bar H_{pol-pol}$, and similarly $\mathcal C(t)=e^{i\bar H_{ph}t} \mathcal C(0)e^{-i\bar H_{ph}t}$ with matrix elements being $\mathcal C(t)_{\vec{k}\lambda,\vec{k'}\lambda'}=\sum_\alpha g^{\lambda\lambda'}_{\alpha}(e^{i\Omega_\alpha t}D^\dagger_{\vec{k-k'},\alpha}-e^{-i\Omega_\alpha t}D_{\vec{k'-k},\alpha})$. In polaron picture, the polaronic and phononic operators separate, and the thermal averages can be performed independently. A detailed derivation of the calculation is provided in Appendix~\ref{ap_gf2}. The resulting spectral function consists of three distinct contributions $\alpha(\omega,T)  =\sum_{k=0}^{2}\alpha^{(k)}(\omega,T)$, with
\begin{subequations}\label{eq:Absorption_T_model2_alpha012}
\begin{align}
  \alpha^{(0)}(\omega,T)
  &= \mathcal W_0(g)
  \sum_{\nu}\sum_{m\in\mathbb{Z}}
  \mathcal L_{m,\nu}(\omega,T,0)
  \label{eq:Absorption_alpha0}
  \\
  \alpha^{(1)}(\omega,T)
  &= \mathcal W_1(g)
  \sum_{\nu}\sum_{m\in\mathbb{Z}}
  (-1)^m \mathcal L_{m,\nu}(\omega,T,|g|)
  \label{eq:Absorption_alpha1}
  \\
  \alpha^{(2)}(\omega,T)
  &= \mathcal W_2(g)
  \sum_{\nu}\sum_{m\in\mathbb{Z}}
  \mathcal L_{m,\nu}(\omega,T,|g|)
  \label{eq:Absorption_alpha2}
\end{align}
\end{subequations}
where the line-shape function is given by
\begin{align}
\mathcal L_{m,\nu}(\omega,T,|g|)
&\equiv
e^{-4\tilde{G}_g(T)}
\frac{e^{\frac12 m\beta\Omega}\,|\psi_\nu(\mathbf 0)|^2\,\gamma}
{(\omega-m\Omega-E_{\mathbf 0\nu})^2+\gamma^2}\nonumber\\
&\qquad I_m\!\left(\frac{4|g|^2}{\sinh(\beta\Omega/2)}\right),
\label{eq:Lmn_def}
\end{align}
and the weight functions are
\begin{align}
\mathcal W_0(g)
&=4\Big[c^4s^4+\big(c^6s^2+c^2s^6\big)e^{-2\tilde G_g(T)}\Big],\nonumber\\
\mathcal W_1(g)
&=2c^4s^4,\qquad
\mathcal W_2(g)=c^8+s^8,
\end{align}
$c=\cos(\theta/2)$, $s=\sin(\theta/2)$ and  $\theta=\arcsin (g_1/|g|)$, $|g|=\sqrt{g_1^2+g_3^2}$, $\tilde{G}_g(T)=[2n_{B}(\Omega,T)+1]|g|^2$. 
The weight functions are plotted in Fig.~\ref{fig:Weight_functions} for different values of $\tilde G_g$.
For small mixing angles $\theta$, i.e. $g_1\ll g_3$, the dominant contribution comes from $\mathcal W_2$.
These weight functions naturally emerge from the structure of the electron-phonon coupling, incorporating strength ratio of intraband and interband couplings via the mixing angle $\theta$. In the diagonal limit $\theta=0$, only the last term $k=2$ contributes, which reduces to Eq.~(\ref{eq:Absorption_T}) as expected.

\section{Results}
We now fit two models to experimental data~\cite{PhononData} for the $\Gamma\rightarrow K$ and $K\rightarrow K$ transitions, labeled by subscripts $\Gamma K$ and $K K$, respectively. For simplicity, we assume a single vibrational mode with momentum-independent Fröhlich coupling and flat phonon dispersion. We also include only a single Wannier exciton state, which is justified by the fact that the typical energy spacing between Wannier levels exceeds the spectral bandwidth of interest.

\begin{figure}
  \centering
  \includegraphics[width=\linewidth]{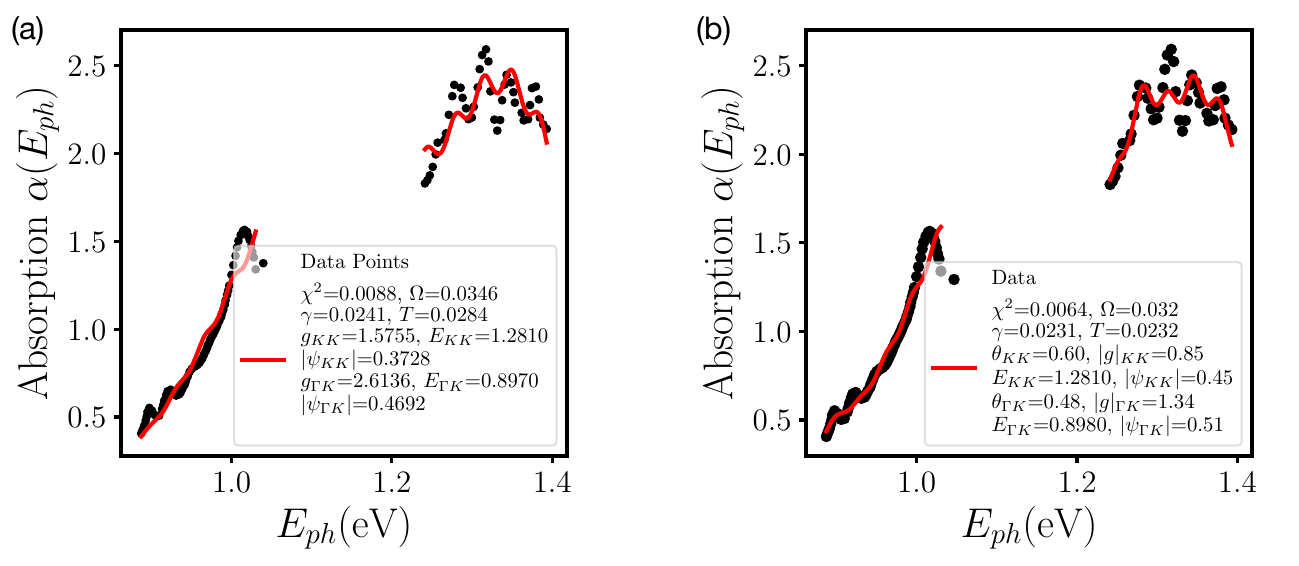}
  \caption{
    Comparison of fits to the experimental data for phonon-assisted absorption models:
    \textbf{(a)} Model I with fixed energy levels, $E_{KK} = 1.281$ eV and $E_{\Gamma K} = 0.897$ eV.
    \textbf{(b)} Model II with $KK$ and $\Gamma K$ interlayer excitons (IX) with fixed energy levels $E_{KK} = 1.281$ eV and $E_{\Gamma K} = 0.898$ eV.
    Here, $\chi^2=N^{-1}\sum_{i=1}^N\left(y_i^{\rm data}-y_i^{\rm fit}\right)^2$ denotes the mean squared fitting error.
  }
  \label{fig:combined}
\end{figure}

\begin{table}
\centering
\caption{
Fitting parameters for Model I and Model II. The exciton energies are fixed during the fitting procedure. In Model I, $|g|$ denotes $g_{cv}=g^c-g^v$; in Model II, $|g|=\sqrt{g_1^2+g_3^2}$ and $\theta$ parametrizes the ratio of off-diagonal to diagonal electron--phonon coupling.
}
\begin{tabular}{lcc}
\toprule
Parameter & Model I & Model II \\
\midrule
$\chi^2$ & 0.0088 & 0.0064 \\
$\Omega$ (eV) & 0.0346 & 0.0320 \\
$\gamma$ (eV) & 0.0241 & 0.0231 \\
$T_{\rm fit}$ (eV) & 0.0284 & 0.0232 \\
$T_{\rm fit}$ (K) & 330 & 270 \\
$E_{KK}$ (eV) & 1.2810 & 1.2810 \\
$E_{\Gamma K}$ (eV) & 0.8970 & 0.8980 \\
$|g|_{KK}$ & 1.5755 & 0.85 \\
$|g|_{\Gamma K}$ & 2.6136 & 1.34 \\
$|\psi_{KK}|$ & 0.3728 & 0.45 \\
$|\psi_{\Gamma K}|$ & 0.4692 & 0.51 \\
$\theta_{KK}$ & -- & 0.60 \\
$\theta_{\Gamma K}$ & -- & 0.48 \\
\bottomrule
\end{tabular}
\label{tab:fitting_parameters}
\end{table}

The fits to both transitions are shown in Fig.~\ref{fig:combined}(a) for Model I and Fig.~\ref{fig:combined}(b) for Model II, with the fitted parameters summarized in Table~\ref{tab:fitting_parameters}. We perform fits with the exciton energy fixed at the positions of two experimentally observed peaks (0.9 eV and 1.3 eV, respectively, as reported in [\onlinecite{PhysRevLett.123.247402}]). It is worth noting that the energy window between $\sim$ 1.0 eV and $\sim$ 1.2 eV is not experimentally accessible currently. As such the experimental data at the edges of this window are also susceptible to errors. The inclusion of interband transition leads to better agreement with the data. This is reflected both in the qualitative similarity and the improved $\chi^{2}$. However, the extracted temperatures ($T_{\rm fit} = 28.4$~meV $\approx 330$~K for Model I, $T_{\rm fit} = 23.2$~meV $\approx 270$~K for Model II) are notably higher than the reported sample temperature of 20~K (1.7~meV)~\cite{PhononData}.
This discrepancy likely reflects local heating effects induced by focused laser heating, which are not captured by the ambient lattice temperature. Transport measurements at room temperature also show signatures of side-band oscillations~\cite{PhononData} suggesting that an elevated temperature does not mitigate the formation of the vibronic phenomena.

Comparing the fits in Fig.~\ref{fig:combined}, Model II provides a better quantitative agreement with experiment ($\chi^2 = 0.0064$) than Model I ($\chi^2 = 0.0088$), despite both capturing most of the main vibronic features. The extracted phonon energy $\Omega$ and linewidth $\gamma$ are similar in both cases, with Model II yielding slightly lower values ($\Omega = 0.032$ eV, $\gamma = 0.0231$ eV) compared to Model I ($\Omega = 0.0346$ eV, $\gamma = 0.0241$ eV). A key distinction lies in the coupling structure. Model I treats the $K\rightarrow K$ and $\Gamma\rightarrow K$ exciton channels as independent, with phonons coupling only to intraband transitions; the coupling strengths $|g|_{KK}$ and $|g|_{\Gamma K}$ are fitted separately for each channel. Model II retains this channel independence but allows phonons to mediate both intraband and interband transitions within each channel, parametrized by an additional mixing angle $\theta$ with $\tan\theta = g_1/g_3$ that controls the ratio of interband to intraband coupling. This produces two additional spectral contributions $\alpha^{(0)}$ and $\alpha^{(1)}$ in Eq.~(\ref{eq:Absorption_T_model2_alpha012}) and modifies the weight of the diagonal contribution $\alpha^{(2)}$ through the weight function $\mathcal{W}_2$, redistributing the intensities of the vibronic sidebands and improving the fit to the experimental peak structure. The improved fit in Model II demonstrates that interband phonon coupling is essential for reproducing the observed sideband structure.

\section{Discussion}

Polaronic effects provide a route to control the flow of energy in optoelectronic devices, both through the renormalization of the effective band structure and through new pathways for phonon-assisted absorption. In this work, we have developed a nonperturbative framework for phonon-assisted optical absorption in TMD heterostructures, based on the polaron transformation, and derived closed-form expressions for the absorption spectrum at arbitrary temperature in two limiting cases: Model 1, restricted to intraband electron-phonon coupling [Eq.~(\ref{eq:Absorption_T})], and Model 2, which additionally includes interband phonon-mediated transitions between the conduction and valence bands [Eq.~(\ref{eq:Absorption_T_model2_alpha012})]. Our analysis shows that interband phonon coupling is essential to reproduce the observed sideband structure, and the mixing angles $\theta_{KK}$ and $\theta_{\Gamma K}$ provide direct measures of the relative strength of interband and intraband phonon-mediated transitions.

While the closed-form expressions assume flat electronic bands, momentum-independent electron-phonon coupling, and a single phonon mode, the general polarization formulas [Eqs.~(\ref{eq:P_model1}) and~(\ref{eq:P_model2})] remain valid for systems with finite phonon dispersion, momentum-dependent coupling, and multiple phonon modes, and can be evaluated numerically in those settings. The framework presented here thus provides a flexible nonperturbative tool for modeling phonon-assisted optical responses in van der Waals heterostructures and other strongly coupled exciton-phonon systems.

\section*{Acknowledgments}
We acknowledge the support  Army Research Office MURI grant no. W911NF-24-1-0292 and Army Research Office Electronic Division award no.W911NF-21-1-0260.






\appendix
\onecolumngrid
\section{Model I: Phonon-assisted intraband absorption}
\subsection{Derivation of the transformed Hamiltonian}\label{ap_mod1}
We derive the Hamiltonian under the unitary transformation. The original Hamiltonian reads $H=H^0_e+H^0_{ph}+H^0_{em}+H_{e-e}+H_{e-ph}+H_{e-em}$ with
\begin{align}
  H^0&=\sum_{\vec k\lambda} \epsilon_{\vec k \lambda} a_{\vec k\lambda}^\dagger a_{\vec k\lambda}
  +\sum_{\vec q\alpha} \Omega_{\vec q\alpha} D_{\vec q\alpha}^\dagger D_{\vec q\alpha}
  +\sum_{\vec q} \omega_{\vec q } B_{\vec q}^\dagger B_{\vec q}\\
  H_{e-e}&=\frac{1}{2}\sum_{\vec {kk'q}\lambda\lambda'}V_{\vec q}a^\dagger_{\vec{k-q}\lambda}
  a^\dagger_{\vec{k'+q}\lambda'}a_{\vec k'\lambda'}a_{\vec k\lambda}\\
  H_{e-ph}&=\sum_{\vec {kq} \lambda\alpha}\Omega_{\vec q\alpha}g_{\vec q\lambda\alpha}
  a^\dagger_{\vec {k-q},\lambda}a_{\vec {k}\lambda}(D_{-\vec q,\alpha}+D^\dagger_{\vec q\alpha})\\
  H_{e-em}&=-\sum_{\vec {kq}\lambda}
           i\mathcal{F}^{\lambda\bar\lambda}_{\vec q} 
           a^\dagger_{\vec {k+q}\lambda}a_{\vec {k}\bar\lambda}B_{\vec q}+h.c..
\end{align}
Here $a_{\vec k\lambda}$, $D_{\vec q\alpha}$ and $B_{\vec q}$  are the annihilation operators of electrons, phonons and photons, respectively. $\lambda$ is the band index referring to the conduction and valence bands, i.e. $\lambda\in\{c,v\}$. $\alpha$ labels the phonon modes. Since the relevant phonon modes vary very little in momentum space, we make the approximation $\Omega_{\vec q\alpha}\approx \Omega_\alpha$. $V_{\vec q}$, $\mathcal{F}^{\lambda\bar\lambda}_{\vec q}$ and $g_{\vec q \lambda\alpha}$ are coupling functions, where $g_{\vec q \lambda\alpha}=g^*_{-\vec q \lambda\alpha}$ to ensure hermiticity of $H_{e-ph}$, and $\mathcal{F}^{cv}_{\vec q}=\mathcal{F}^{vc*}_{\vec q}=\mathcal{F}_{\vec q}$.

The idea of polaron transformation is to perform a canonical transformation $U=e^{S}$ so that the $H_{e-eh}$ in the transformed Hamiltonian $e^{\textup{ad}_S}H$ vanishes, while remaining terms are renormalized. To eliminate ${e-eh}$, we need to find an anti-Hermitian operator $S$ so that 
\begin{align}
  \textup{ad}_S H^0_{ph}+H_{e-ph}=0
\end{align}
with
\begin{align}
  S
  &=\sum_{\vec{kq}\lambda\alpha}g_{\vec q\lambda\alpha}
  (D^\dagger_{\vec q\alpha}-D_{-\vec q,\alpha})a_{\vec {k-q,\lambda}}^\dagger a_{\vec k\lambda},
  =\sum_{\vec{kq}\lambda\alpha}g_{\vec q\lambda\alpha}Q_{\vec q\alpha}
  a_{\vec {k-q,\lambda}}^\dagger a_{\vec k\lambda},
\end{align}
and $Q_{\vec q\alpha}=D^\dagger_{\vec q\alpha}-D_{-\vec q,\alpha}$.

We now derive the transformed Hamiltonian. For later convenience we first derive the adjoin action of $S$ on the creation/annihilation operators of electrons, phonons and photons.
\begin{itemize}
  \item  The action of $\textup{ad}_S$ on $a_{\vec k\lambda}$ is given by 
  \begin{align}
    \textup{ad}_S a_{\vec k\lambda}
    &=-\sum_{\vec {p}\alpha}g^\lambda_{\vec{p-k},\alpha}Q_{\vec{p-k},\alpha}a_{\vec p\lambda}
    =-\sum_{\vec p}a_{\vec p\lambda}\mathcal C^\lambda_\vec{p,k}\\
    \textup{ad}_S a^\dagger_{\vec k\lambda}
    &=\sum_{\vec {p}\alpha}g^\lambda_{\vec{k-p},\alpha}Q_{\vec{k-p},\alpha}a^\dagger_{\vec p\lambda}
    =\sum_{\vec p}\mathcal C^\lambda_\vec{k,p}a^\dagger_{\vec p\lambda}
  \end{align}
  where we have defined $\mathcal C^\lambda_\vec{k,k'}\equiv \sum_\alpha g^\lambda_{\vec{k-k'},\alpha}
  Q_{\vec{k-k'},\alpha}$ as the matrix element of the anti-Hermitian operator $\mathcal C^\lambda$. 
  This allows to write the adjoin action of $U=e^S$ on electronic operators as
  \begin{align}
    \textup{Ad}_Ua_{\vec k\lambda}&=e^{\textup{ad}_S} a_{\vec k\lambda}
    =\sum_{\vec p}a_{\vec p\lambda}(e^{-\mathcal{C}^\lambda})_{\vec{p,k}}\label{eq:a}\\
    \textup{Ad}_Ua^\dagger_{\vec k\lambda}&=e^{\textup{ad}_S} a^\dagger_{\vec k\lambda}
    =\sum_{\vec p}(e^{\mathcal{C}^\lambda})_{\vec{k,p}}a^\dagger_{\vec p\lambda}\label{eq:a_dagger}
  \end{align}
  \item  The action of $\textup{ad}_S$ on $D_{\vec q\alpha}$ is 
  \begin{align}
    \textup{ad}_S D_{\vec q\alpha} &=-\sum_{\vec{k}\lambda}
     g_{\vec q\lambda\alpha} a_{\vec {k-q,\lambda}}^\dagger a_{\vec k\lambda}\\
     \textup{ad}_S D^\dagger_{\vec q\alpha} &=-\sum_{\vec{k}\lambda}
     g^*_{\vec q\lambda\alpha} a_{\vec k \lambda}^\dagger a_{\vec{k-q}\lambda},
  \end{align}
  which yields $\textup{ad}_S D^\dagger_{\vec q\alpha}D_{\vec q\alpha} =-\sum_{\vec{k}\lambda}
   g_{\vec q\lambda} a_{\vec {k-q,\lambda}}^\dagger a_{\vec k\lambda}
   (D^\dagger_{\vec q \alpha}+D_{-\vec q,\alpha})$, and thus $\textup{ad}_SH^0_{ph}+H_{e-ph}=0$ as promised.
  \item  Since $S$ commutes with $B_{\vec q}$, then
  \begin{align}
    \textup{ad}_S B_{\vec q} =0,\qquad e^{\textup{ad}_S} B_{\vec q} =0.
  \end{align}
\end{itemize}
The transformed Hamiltonian is obtained as follows:
\begin{itemize}
  \item[(1)] Defining the matrix element 
  $\mathcal{E}^\lambda_{\vec {kk'}}=\epsilon_{\vec k\lambda}\delta_{\vec {kk'}}$  Eq.(\ref{eq:a}) and Eq.(\ref{eq:a_dagger}) lead to 
  \begin{align} 
    e^{\textup{ad}_S} H^0_e
    &=\sum_{\vec k\lambda} \epsilon_{\vec k \lambda} 
     (e^{\textup{ad}_S}a_{\vec k\lambda}^\dagger) (e^{\textup{ad}_S}a_{\vec k\lambda})\nonumber\\
    &=\sum_{\vec {pp'k}\lambda} \epsilon_{\vec k \lambda} 
    (e^{\mathcal{C}^\lambda})_{\vec{k,p}}a^\dagger_{\vec p\lambda}
    a_{\vec p'\lambda}(e^{-\mathcal{C}^\lambda})_{\vec{p',k}}\nonumber\\
    &=\sum_{\vec {pp'k}\lambda}(e^{-\mathcal{C}^\lambda}\mathcal{E}^\lambda e^{\mathcal{C}^\lambda})_\vec{p'p}
    a^\dagger_{\vec p\lambda}a_{\vec p'\lambda}.
  \end{align}
  \item[(2)] $\textup{ad}_S H_{e-ph}$ is given by
  \begin{align}
    \textup{ad}_S H_{e-ph}
    &=\textup{ad}_S \sum_{\vec {kq} \lambda\alpha}\Omega_{\vec q\alpha}g_{\vec q\lambda\alpha}
    a^\dagger_{\vec {k-q},\lambda}a_{\vec {k}\lambda}(D_{-\vec q,\alpha}+D^\dagger_{\vec q\alpha})\nonumber\\
    &=-2\sum_{\vec{kk'q}\lambda\lambda'\alpha} \Omega_{\vec q\alpha}g_{\vec q\lambda\alpha}g^*_{\vec q\lambda'\alpha}
    a^\dagger_{\vec{k-q}\lambda}a_{\vec k\lambda}
    a^\dagger_{\vec{k'+q}\lambda'}a_{\vec k'\lambda'}\nonumber \\
    &=-2\sum_{\vec{kq}\lambda\alpha} \Omega_{\vec q\alpha}|g_{\vec q\lambda\alpha}|^2
    a^\dagger_{\vec{k}\lambda}a_{\vec {k}\lambda}
    -2\sum_{\vec{kk'q}\lambda\lambda'\alpha} (\Omega_{\vec q\alpha}g_{\vec q\lambda\alpha}g^*_{\vec q\lambda'\alpha}) 
    a^\dagger_{\vec{k-q}\lambda}a^\dagger_{\vec{k'+q}\lambda'}a_{\vec k'\lambda'}a_{\vec k\lambda}.
  \end{align}
  An useful identity is
  \begin{align}
    (\textup{ad}_S)^2 H_{e-ph}=0 \qquad ((\textup{ad}_S)^n H_{e-ph}=0, \textup{ for }\forall n\ge2)
  \end{align} 
  This is derived by rewriting $S$ and $\textup{ad}_S H_{e-ph}$ in real space. Notice that 
  \begin{align*}
    \textup{ad}_S H_{e-ph}
    &=-2\sum_{\lambda\lambda'\alpha}\int_{\vec {r,r'}}
    [\Omega_\alpha\star g_{\lambda'\alpha}\star g^*_{\lambda\alpha}](\vec r-\vec r')
    \rho_\lambda({\vec r})\rho_{\lambda'}({\vec r'})\\
    S&=\sum_{\lambda\alpha}\int_{\vec r}[g_{\lambda\alpha}\star Q_\alpha](\vec r)\rho_{\lambda}(\vec r),
  \end{align*}
  where $[\Omega_\alpha\star g_{\lambda'\alpha}\star g^*_{\lambda\alpha}]$ and $[g_{\lambda\alpha}\star Q_\alpha]$ are the convolutions of the real space representation of each term inside, $\rho_\lambda({\vec r})$ is the spatial density operator of electron in band $\lambda$. The commutation relation of density operator $[\rho_\lambda({\vec r}),\rho_{\lambda'}({\vec r'})]=0$ yields $(\textup{ad}_S)^2 H_{e-ph}=0$.
  \item[(3)] The renormalization of $H^0_{ph}$ and $H_{e-ph}$:
  \begin{align} 
    e^{\textup{ad}_S} H^0_{ph}+e^{\textup{ad}_S} H_{e-ph}
    &=H^0_{ph}+\sum_{n=1}^\infty\frac{1}{n!}(\textup{ad}_S)^n H^0_{ph}
    +\sum_{n=0}^\infty\frac{1}{n!}(\textup{ad}_S)^n H_{e-ph}\nonumber \\
    &=H^0_{ph}+\sum_{n=1}^\infty\frac{n}{(n+1)!}(\textup{ad}_S)^n H_{e-ph}\nonumber\\
    &=H^0_{ph}+\frac{1}{2}\textup{ad}_S H_{e-ph},
  \end{align}
  where we have used the relation $ \textup{ad}_S H^0_{ph}=-H_{e-ph}$, and the fact that 
  $(\textup{ad}_S)^n H_{e-ph}=0, \textup{ for }\forall n\ge2$.
  \item[(4)] The free photon Hamiltonian remains the same: 
  \begin{align} 
    e^{\textup{ad}_S} H^0_{em}=H^0_{em}.
  \end{align}
  \item[(5)] The renormalized photon-electron interaction is 
  \begin{align} 
    e^{\textup{ad}_S} H_{e-e}
    &=\frac{1}{2}\sum_{\vec {kk'q}\{\vec k_i\}\lambda\lambda'}V_{\vec q}
    (e^{\mathcal{C}^\lambda})_{\vec{k-q,k}_1} 
    (e^{\mathcal{C}^{\lambda'}})_{\vec{k'+q,k}_2} 
    (e^{-\mathcal{C}^{\lambda'}})_{\vec k_3,\vec k'} 
    (e^{-\mathcal{C}^\lambda})_{\vec k_4,\vec k} 
    \nonumber \\
    &\qquad \qquad a^\dagger_{{\vec k_1}\lambda}a^\dagger_{\vec k_2 \lambda'}a_{\vec k_3\lambda'}a_{\vec k_4\lambda}.
  \end{align}
  Notice that $\sum_{\vec k'} (e^{\mathcal{C}^{\lambda'}})_{\vec{k'+q,k}_2} (e^{-\mathcal{C}^{\lambda'}})_{\vec k_3,\vec{k'}}=\delta_{\vec{k}_2-\vec q,\vec k3}$ and similarly $\sum_{\vec k} (e^{\mathcal{C}^{\lambda}})_{\vec{k-q,k}_1} (e^{-\mathcal{C}^{\lambda}})_{\vec k_4,\vec{k}}=\delta_{\vec{k}_1+\vec q,\vec k4}$, which directly follow from the property $\mathcal{C}^{\lambda}_{\vec k\vec k'}=\mathcal{C}^{\lambda}_{\vec {k+q},\vec {k'+q}}$. Therefore, the Coulomb interaction term is not renormalized by $S$,
  \begin{align}
    e^{\textup{ad}_S} H_{e-e}= H_{e-e}.
  \end{align}
  \item[(6)] We finally calculate the renormalized electron-photon interaction, which is given by 
  \begin{align}
    e^{\textup{ad}_S} H_{e-em}
   &= -\sum_{\vec {kq}\{\vec k_i\}\lambda }
    i\mathcal{F}^{\lambda\bar\lambda}_{\vec q} 
    (e^{\mathcal{C}^\lambda})_{\vec{k+q,k}_1} 
    (e^{-\mathcal{C}^{\bar\lambda}})_{\vec k_2 \vec k}
    a^\dagger_{\vec {k}_1\lambda}
    a_{\vec {k}_2\bar\lambda}B_{\vec q}+h.c. \nonumber \\
   &= -\sum_{\vec k_1\vec k_2 \vec q \lambda}
    i\mathcal{F}^{\lambda\bar\lambda}_{\vec q} 
    (e^{\mathcal{C}^\lambda-\mathcal{C}^{\bar\lambda}})_{\vec k_2+\vec q,\vec k_1} 
    a^\dagger_{\vec {k}_1\lambda}
    a_{\vec {k}_2\bar\lambda}B_{\vec q}+h.c.. 
  \end{align} 
\end{itemize}
In summary, the transformed total Hamiltonian reads 
\begin{align}
\bar H_{pol}
  &=\sum_{\vec {pp'}\lambda}(e^{-\mathcal{C}^\lambda}\mathcal{E}^\lambda e^{\mathcal{C}^\lambda})_\vec{p'p}
  a^\dagger_{\vec p\lambda}a_{\vec p'\lambda}-\sum_{\vec{kq}\lambda\alpha} \Omega_{\alpha}|g_{\vec q\lambda\alpha}|^2
  a^\dagger_{\vec{k}\lambda}a_{\vec {k}\lambda} \\
\bar H^0_{ph}
  &=\sum_{\vec q\alpha}\Omega_{\alpha} D^\dagger_{\vec q\alpha}D_{\vec q\alpha} \qquad\qquad
\bar H^0_{em}
  =\sum_{\vec q}\omega_\vec q B^\dagger_\vec q B_\vec q \\
\bar H_{pol-pol}
  &=\frac{1}{2}\sum_{\vec{kk'q}\lambda\lambda'} 
  (
    V_\vec q-2\sum_{\alpha}\Omega_{\alpha} g_{\vec q\lambda\alpha}g^*_{\vec q\lambda'\alpha}
  )
  a^\dagger_{\vec{k-q}\lambda}a^\dagger_{\vec{k'+q}\lambda'}a_{\vec k'\lambda'}a_{\vec k\lambda} \\
\bar H_{pol-em}
  &=-\sum_{\vec k_1\vec k_2 \vec q \lambda}
  i\mathcal{F}^{\lambda\bar\lambda}_{\vec q} 
  (e^{\mathcal{C}^\lambda-\mathcal{C}^{\bar\lambda}})_{\vec k_2+\vec q,\vec k_1} 
  a^\dagger_{\vec {k}_1\lambda}
  a_{\vec {k}_2\bar\lambda}B_{\vec q}+h.c..  
\end{align}
Here the electron-phonon coupling is removed explicitly by the canonical transformation, but is incorporated in polaron dispersion relation and in the renormalized polaron-photon coupling function with phonon operators involved. The polaron dispersion relation is given by $e_{\vec k\lambda}(T)=e^{-\tilde{G}_\lambda(T)}\sum_{\vec k}\left(e^{G_\lambda(T)}\right)_{\vec{pk}}\epsilon_{\vec k\lambda}-\Omega \tilde{G}_\lambda$, with $\tilde G_\lambda(T)=\sum_{\vec q\alpha}(2N_{\vec q\alpha}+1)|g^\lambda_{\vec q\alpha}|^2$, and $[G_\lambda(T)]_{\vec{kk'}}=\sum_\alpha(2N_{\vec{k-k'},\alpha}+1)|g^\lambda_{\vec {k-k'},\alpha}|^2$.

\subsection{Equations of Motion}\label{ap_eom}

Equation of motion approach is used to determine the absorption spectrum. The approach can be easily adapted to the emission spectrum as well. Normal ordering the phonon operators:
\begin{align}
    \label{eq:EOM_Absorption}
    A(t) &= \sum_{\varepsilon_\lambda > \varepsilon_{\lambda'}}\sum_{\vk,\vq}\mathcal{F}^{\lambda,\lambda'}\sum_{\alpha,\beta} C^{\alpha,\beta}_{\lambda,\lambda'}\Big(BP^{\lambda,\lambda'}_{\vk-\vq,\vk}\Big)\nonumber\\
    &\,\,\qquad\qquad\qquad\qquad\times\Big([d^\dagger_{\lambda,\lambda'}]^\alpha [d^T_{\lambda,\lambda'}]^\beta\Big)_{\vq,0}
\end{align}
Where $\alpha$ ($\beta$) represents the number of emitted (absorbed) phonons. The normal ordering factor and scaled phonon operators are given by:
\begin{align}
    C^{\alpha,\beta}_{\lambda,\lambda'}&\equiv e^{-\tfrac{1}{2}\tilde{G}_{\lambda,\lambda'}}\frac{(-1)^\beta}{(\alpha!)(\beta!)}\\
    \tilde{G}_{\lambda,\lambda'}&\equiv\sum_{\nu,\vq}|g^\lambda_{\nu,\vq}-g^{\lambda'}_{\nu,\vq}|^2\\
    (d^\dagger_{\lambda,\lambda'})_{\vk,\vk'}&\equiv\sum_\nu (g^\lambda_{\nu,\vk-\vk'}-g^{\lambda'}_{\nu,\vk-\vk'})D^\dagger_{\nu,\vk-\vk'}\nonumber\\
    &\equiv \sum_{\nu}[(d^\dagger)^\nu_{\lambda,\lambda'}]_{\vk,\vk'}\\
    (d^T_{\lambda,\lambda'})_{\vk-\vk'}&\equiv(d_{\lambda,\lambda'})_{\vk'-\vk}
\end{align}
For a single dispersionless phonon branch with momentum independent Fr\"ohlich matrix elements, the polarization function defined in Eq.(\ref{eq:EOM_Absorption}) obeys a Wannier equation \cite{Feldtmann}. To account for the more general interaction, we expand the phonon piece of Eq.(\ref{eq:EOM_Absorption}):
\begin{align}
    \label{eq:Phonon_Products}
    \Big([d^\dagger_{\lambda,\lambda'}]^\alpha [d^T_{\lambda,\lambda'}]^\beta\Big)_{\vq,0} &= \sum_{\{\nu_i,\vq'_i\}}\sum_{\{\nu_j,\vq'_j\}}\prod_{i=0}^{\alpha-1}[(d^\dagger)^{\nu_i}_{\lambda,\lambda'}]_{\vq'_i}\prod_{j=\alpha}^{\alpha+\beta-1}[(d^T)^{\nu_j}_{\lambda,\lambda'}]_{\vq'_j}
\end{align}
with the momenta defined as $\vq'_i\equiv\vq_i-\vq_{i+1}$, $\vq_0\equiv\vq$, and $\vq_{\alpha+\beta}=0$, such that $\sum_i\vq'_i=\vq$. We are now equipped to define the multi-particle correlation function for the most general interacting polaron picture:
\begin{align}
    \label{eq:Correlation_Function}
    \bar{\Pi}^{\lambda,\lambda'}_{\alpha\beta}(\vk,\vq) &\equiv \expval{\Pi^{\lambda,\lambda'}_{\alpha\beta}(\vk,\vq)} = \expval{BP^{\lambda,\lambda'}_{\vk-\vq,\vk}\prod_{i=0}^{\alpha-1}[(d^\dagger)^{\nu_i}_{\lambda,\lambda'}]_{\vq'_i}\prod_{j=\alpha}^{\alpha+\beta-1}[(d^T)^{\nu_j}_{\lambda,\lambda'}]_{\vq'_j}}
\end{align}
The absorption contributions to the average photon occupation number become:
\begin{align}
    \label{eq:Absorption_Correlation}
    \expval{A(t)} &= \sum_{\varepsilon_\lambda > \varepsilon_{\lambda'}}\sum_{\vk,\vq}\mathcal{F}^{\lambda,\lambda'}\sum_{\alpha,\beta} C^{\alpha,\beta}_{\lambda,\lambda'}\sum_{\{\nu_i,\vq'_i\}}\sum_{\{\nu_j,\vq'_j\}}\bar{\Pi}^{\lambda,\lambda'}_{\alpha\beta}(\vk,\vq)
\end{align}
To obtain the absorption spectrum given by Eq.(\ref{eq:Absorption_Correlation}), we once again apply the equation of motion approach, this time to the correlation function defined in Eq.(\ref{eq:Correlation_Function}). We first define an electronic band that accounts for the Coulomb self energy:
\begin{align}
    \tilde{e}_{\lambda,\vk} &= e_{\lambda,\vk} - \sum_{\vk'}\tilde{V}^{\lambda,\lambda}_{\vk-\vk'}f^\lambda_{\vk'}
\end{align}
where $f^\lambda_{\vk}$ is the Fermi distribution. We consider the equation of motion of Eq.(\ref{eq:Correlation_Function}) for an arbitrary conduction band $\lambda=c$ and valence band $\lambda'=v$, under the initial conditions $f^v_\vk = 1$ and $f^c_\vk=0$ for arbitrary $\vk$. Under these conditions, the correlation function satisfies the following equation of motion to zeroth order in the cluster expansion:
\begin{align}
    \label{eq:Corr_EOM}
    &\bigg\{-i\hbar\partial_t + (\tilde{e}_{c,\vk-\vq}-\tilde{e}_{v,\vk}) + \sum_{i=0}^{\alpha-1}(\bar\Omega^{\nu_i}_{\vq'_i})-\sum_{j=0}^{\alpha+\beta-1}(\hbar\Omega^{\nu_j}_{\vq'_j})-\hbar\omega\bigg\}\bar{\Pi}^{c,v}_{\alpha\beta}(\vk,\vq)\nonumber\\
    &\qquad= \sum_{\vk'}\tilde{V}^{v,c}_{\vk-\vk'}\bar{\Pi}^{c,v}_{\alpha\beta}(\vk',\vq) + iS^{c,v}_{\alpha,\beta}(\vk,\vq,t)
\end{align}
The source term in Eq.(\ref{eq:Corr_EOM}) is given by:
\begin{align}
    \label{eq:Source}
    S^{c,v}_{\alpha,\beta}(\vk,\vq,t) &= \mathcal{F}^{v,c}\expval{B^\dagger B}\expval{(e^{C_v-C_c})_{-\vq,0}\prod_{i=0}^{\alpha-1}[(d^\dagger)^{\nu_i}_{c,v}]_{\vq'_i}\prod_{j=\alpha}^{\alpha+\beta-1}[(d^T)^{\nu_j}_{c,v}]_{\vq'_j}}
\end{align}
We recognize Eq.(\ref{eq:Corr_EOM}) as the usual Wannier Equation \cite{doi:10.1142/7184}, now modified to account for phonon interactions. We solve it in the usual manner by expanding in a basis of Wannier wave functions. The absorption coefficient $\alpha(\omega)$ defined in terms of the absorption spectrum is:
\begin{align}
    \alpha(\omega)\equiv \Re{A(\omega,t)/\expval{B^\dagger B}}
\end{align}
This yields the following Lorentzian expression for the absorption coefficient:
\begin{align}
    \label{eq:Absorption_Zero}
    \alpha(\omega) &= \sum_{\varepsilon_c>\varepsilon_v}\sum_{\alpha,\beta}\sum_{\vq}C^{\alpha,\beta}_{c,v}|F^{c,v}|^2\sum_{\{\alpha,\beta\}}\sum_{n}\frac{D^{c,v}_{\{\alpha,\beta\}}(\vq,T)\,|\Psi^{c,v}_{n,\{\alpha,\beta\}}(0)|^2\,\Gamma^{c,v}_{\{\alpha,\beta\}}}{\big(\tilde{E}^{\{\alpha,\beta\}}_{g,\vq} + \Delta E^{\{\alpha,\beta\}}_{ph}-E^{\{\alpha,\beta\}}_n-\hbar\omega\big)^2+\big(\Gamma^{c,v}_{\{\alpha,\beta\}}\big)^2}
\end{align}
where we have introduced the notation $\{\alpha,\beta\}$ to represent a configuration of the phonon branch and momentum of the absorbed and emitted phonons. Eq.(\ref{eq:Absorption_Zero}) includes a sum over all Wannier energies $E^{\{\alpha,\beta\}}_n$ with wave function $\Psi^{c,v}_{n,\{\alpha,\beta\}}(\mathbf{r})$. The modified gap energy $\tilde{E}^{\{\alpha,\beta\}}_{g,\vq}$ accounts for both the modified band gap of the polaron band structure as well as an associated kinetic energy from the exciton center of mass. $\Delta E^{\{\alpha,\beta\}}_{ph}$ represents the net energy gained (lost) from net absorption (emission) of phonons. The broadening coefficient $\Gamma^{c,v}_{\{\alpha,\beta\}}$ accounts for the exciton lifetime. 
\par The factor $D^{c,v}_{\{\alpha,\beta\}}(\vq,T)$ in Eq.(\ref{eq:Absorption_Zero}) are the expectation value of the phonon operators of Eq.(\ref{eq:Source}) at temperature $T$, and is related to the overall amplitude of a given phonon configuration side band. At finite temperature, it is given by:
\begin{align}
    \label{eq:EOM_Amplitude}
    D^{c,v}_{\{\alpha,\beta\}}(\vq,T) &= \frac{1}{\alpha ! \beta !}e^{-\bar{G}_{c,v}-\tfrac{1}{2}\tilde{G}_{c,v}}\prod_{i=0}^{\alpha-1}(\bar{G}_{c,v}+\mathbb{1})_{\nu_i,\vq'_i}\prod_{j=\alpha}^{\alpha+\beta-1}(\bar{G}_{c,v})_{\nu_l,\vq'_j}
\end{align}
Where $(\bar{G}_{c,v})_{\vk,\vk'}\equiv\sum_{\nu}|g^c_{\nu\vk-\vk'}-g^v_{\nu\vk-\vk'}|^2n_{\nu,\vk-\vk'}$ and $n_{\nu,\vk-\vk'}$ is the thermal averaged phonon occupation number. 

\subsection{Green's Function}\label{ap_gf}
\subsubsection{Linear response theory revisited}
The optical-absorption and emission coefficient is described by the rate of change of photon number operator
\cite{segall1968phonon}, namely
\begin{align}\label{eq:intensity}
  I(\omega_\vec q)=\partial_t\langle B^\dagger_\vec q B_\vec q\rangle
  \sim -[\alpha(\omega_\vec q)-e(\omega_\vec q)] \langle B^\dagger_\vec q B_\vec q\rangle,
\end{align}
where the proportionality constant between the photon flux $I(\omega_\vec q)$ and the photon numbers $\langle B^\dagger_\vec q B_\vec q\rangle$ is identified as the difference between absorption and emission coefficients.  The electron and hole distributions are defined as $f_e=n^F_c=\langle a^\dagger_c a_c\rangle$ and $f_h=1-n^F_c=1-\langle a^\dagger_v a_v\rangle$, respectively. The absorption coefficient is proportional to the joint probability of having unoccupied electron and hole states, i.e. $\alpha(\omega_\vec q)\propto (1-f_e)(1-f_h)$ and the emission $e(\omega_\vec q)\propto f_ef_h$.

Let $H=H^0_e+H^0_{ph}+H^0_{em}+H_{e-e}+H_{e-ph}$ be the unperturbed Hamiltonian. In the following, we will treat the light-matter coupling term $H_{e-em}$ as the external perturbation. For later convenience, we rewrite $H_{e-em}$ as
\begin{align}
  H_{e-em}=i\sum_{\vec q\lambda}(\mathcal{F}_{\vec q}^{\lambda\bar\lambda}P^{\lambda\bar\lambda}_\vec qB_{\vec q}
           -\mathcal{F}_{\vec q}^{\bar\lambda\lambda}P^{\bar\lambda\lambda}_\vec{-q}B^\dagger_{\vec q}),
\end{align}
where we have defined the polarization operator as $P^{\lambda\bar\lambda}_\vec q=\sum_{\vec k}a^\dagger_{\vec {k+q}\lambda}a_{\vec k\bar\lambda}$. To linear order in $H_{e-em}$, the change of $\partial_t(B^\dagger_\vec q B_\vec q)$ evaluated with respect to the ground state of $H$ is given by 
\begin{align}\label{eq:dynamicsBB}
  \langle\partial_t(B^\dagger_\vec q B_\vec q) \rangle
  =-i\int_{-\infty}^t dt'\left\langle[\partial_t(B^\dagger_\vec q B_\vec q)(t),H_{e-em}(t')]\right\rangle\nonumber\\
  =\int_{-\infty}^t dt'\left\langle\left[[H_{e-em}(t),(B^\dagger_\vec q B_\vec q)(t)],H_{e-em}(t')\right]\right\rangle.
\end{align}
Since
\begin{align}
  &[H_{e-em}(t),(B^\dagger_\vec q B_\vec q)(t)]\nonumber\\
  &=i\sum_{\vec q\lambda}[\mathcal{F}_{\vec q}^{\lambda\bar\lambda}P^{\lambda\bar\lambda}_\vec q(t)B_{\vec q}(t)
  +\mathcal{F}_{\vec q}^{\bar\lambda\lambda}P^{\bar\lambda\lambda}_\vec{-q}(t)B^\dagger_{\vec q}(t)].
\end{align}
we have
\begin{align}\label{eq:HBBH}
  &\left\langle\left[[H_{e-em}(t),(B^\dagger_\vec q B_\vec q)(t)],H_{e-em}(t')\right]\right\rangle
   \nonumber\\
  &
  \sim \langle B^\dagger_\vec q B_\vec q \rangle
  \sum_{\lambda\lambda'}\mathcal{F}_{\vec q}^{\lambda\bar\lambda}\mathcal{F}_{\vec q}^{\bar\lambda'\lambda'}
  \left\langle
   \left[
    P^{\lambda\bar\lambda}_\vec{q}(t),P^{\bar\lambda'\lambda'}_\vec{-q}(t')
   \right]
  \right\rangle 
  e^{-i\omega_\vec q(t-t')}\nonumber\\
  &
  -\langle B^\dagger_\vec q B_\vec q \rangle
  \sum_{\lambda\lambda'}\mathcal{F}_{\vec q}^{\bar\lambda\lambda}\mathcal{F}_{\vec q}^{\lambda'\bar\lambda'}
  \left\langle
   \left[
    P^{\bar\lambda\lambda}_\vec{-q}(t),P^{\lambda'\bar\lambda'}_\vec{q}(t')
   \right]
  \right\rangle 
  e^{i\omega_\vec q(t-t')},
\end{align}
where we only keep terms that are proportional to the photon occupation number $\langle B^\dagger_\vec q B_\vec q \rangle$. For optical transitions, we can take $\vec q=0$ and define $\omega_{\vec q=0}=\omega$, $\mathcal{F}_{\vec q=0}^{\bar\lambda\lambda}=\mathcal{F}^{\bar\lambda\lambda}$. Inserting Eq.(\ref{eq:HBBH}) into Eq.(\ref{eq:dynamicsBB}), and noticing that Wick theorem constrains that $\lambda=\lambda'$, we obtain 
\begin{align}
  \alpha(\omega)-e(\omega)
  &=-2\sum_{\lambda}|\mathcal{F}^{\lambda\bar\lambda}|^2
  \Im[\mathcal G^R_{ P^{\bar\lambda\lambda}_0,P^{\lambda\bar\lambda}_0}(\omega)]
\end{align}
in which retarded and advanced Green's function of operators $\mathcal O_1$, $\mathcal O_2$ are defined as 
\begin{align}
  \mathcal G^R_{\mathcal O_1,\mathcal O_2}(t)
  &=-i\Theta(t)
  \left\langle
   \left[
    \mathcal O_1(t),\mathcal O_2(0)
   \right]
  \right\rangle \nonumber\\
  \mathcal G^A_{\mathcal O_1,\mathcal O_2}(t)
  &=i\Theta(-t)
  \left\langle
   \left[
    \mathcal O_1(t),\mathcal O_2(0)
   \right]
  \right\rangle. \nonumber
\end{align}
The absorption coefficient consists of two terms, $\Im[\mathcal G^R_{ P^{vc}_0,P^{cv}_0}(\omega)]$ and $\Im[\mathcal G^R_{ P^{cv}_0,P^{vc}_0}(\omega)]$. The first term represents the resonant part which contributes to the absorption, while the second term corresponds to the non-resonant contribution in the sense that the poles in the complex frequency plane lie in the left half plane \cite{haug2009quantum}. Therefore, the second term can be omitted as it is a regular function for $\omega>0$. 

Given that $\alpha(\omega)\propto (1-f_e)(1-f_h)$ and $e(\omega)\propto f_ef_h$, we split the commutator in the retarded Green's function $\mathcal G^R_{ P^{vc}_0,P^{cv}_0}(\omega)$ into greater and lesser components. The imaginary parts of these components give rise to the absorption and emission coefficients, respectively. Rescaling the Green's function by dividing the number of unit cells $N$ in the sample, we have
\begin{align}
  P^R_> (t)
  &=-i\frac{\Theta(t)}{N}\sum_{\vec k\vec k'}
    \langle
    a^\dagger_{\vec k v}(t) a_{\vec k c}(t) a^\dagger_{\vec k' c}(0) a_{\vec k' v} (0)
    \rangle 
  \\
  \alpha(\omega)
  &\propto- \Im P^R_>(\omega)
  \\
  P^R_<(t)
  &=-i\frac{\Theta(t)}{N}\sum_{\vec k\vec k'}
    \langle
    a^\dagger_{\vec k c}(0) a_{\vec k v}(0) a^\dagger_{\vec k' v}(t) a_{\vec k' c} (t)
    \rangle
  \\
  e(\omega)     
  &\propto- \Im P^R_<(\omega).
\end{align}
The definition of absorption coefficient is consistent with Ref. [\onlinecite{hannewald2005nonperturbative}].

\subsubsection{Absorption spectrum at finite temperatures}
In the following, we calculate the absorption coefficient through the greater part of the retarded four-point correlation function. In linear response theory, we treat the carrier-photon coupling terms as an external perturbation. Thus the time-dependence of electronic operator is governed by the Hamiltonian without $H_{e-em}$ and $H^0_{em}$. However, the coupling of carriers with phonons complicates the calculation of the correlation functions. To circumvent this we resort to the Polaron picture by performing a unitary transformation so that the electron-phonon coupling is incorporated nonperturbatively. From now on, $H$ denotes the full Hamiltonian without $H_{e-em}$ and $H^0_{em}$, i.e. $H=H^0_e+H^0_{ph}+H_{e-e}+H_{e-ph}$, unless otherwise stated. Under the the canonical transformation,
\begin{align}
  Ua^{\dagger}_{\vec k \lambda}(t)U^\dagger
  &=\sum_{\vec p}e^{i(\bar H_{pol}+\bar H_{pol-pol})t}
    a^{\dagger}_{\vec k \lambda}e^{-i(\bar H_{pol}+\bar H_{pol-pol})t} 
    e^{i\bar H^0_{ph}t}(e^{\mathcal C^\lambda})_{\vec{kp}}e^{-i\bar H^0_{ph}t}\nonumber 
\end{align}
we obtain
\begin{align}
  P^R_>(t)
  &=-i\frac{\Theta(t)}{N}\sum_{\vec {k_1...k_4}}
  \langle
  a^\dagger_{\vec k_1 v}(t) a_{\vec k_2 c}(t) a^\dagger_{\vec k_3 c}(0) a_{\vec k_4 v} (0)
  \rangle
  \langle 
  (e^{ \mathcal C^v(t)- \mathcal C^c(t)})_{\vec {k_2k_1}} (e^{ \mathcal C^c(0)- \mathcal C^v(0)})_{\vec {k_4k_3}}
  \rangle,
\end{align}
where polaron operator evolves according to $a^{(\dagger)}_{\vec k \lambda}(t)=e^{i\bar H_{p}t} a^{(\dagger)}_{\vec k \lambda}e^{-i\bar H_{p}t}$ and $\bar H_p=\bar H_{pol}+\bar H_{pol-pol}$, and similarly $ \mathcal C^\lambda(t)=e^{i\bar H_{ph}t}  \mathcal C^\lambda(0)e^{-i\bar H_{ph}t}$ with matrix elements being $  \mathcal C^\lambda(t)_{\vec{kk'}}=\sum_\alpha g^\lambda_{\vec{k-k'},\alpha}(e^{i\Omega_\alpha t}D^\dagger_{\vec{k-k'},\alpha}-e^{-i\Omega_\alpha t}D_{\vec{k'-k},\alpha})$. In polaron picture the polaronic and phononic operators separate so that the thermal averages can be performed independently.

To evaluate the thermal average of phonon operators at finite temperatures, we generalize the method of Feynman disentangling of operators in Ref.[\onlinecite{munn1985theory}], by introducing the tensor product notation 
\begin{align}\label{eq:thavg_ph}
 &\langle 
  (e^{ \mathcal C^v(t)- \mathcal C^c(t)})_{kk'} (e^{ \mathcal C^c(0)-  \mathcal C^v(0)})_{qq'}
  \rangle\nonumber\\
  &\qquad=\langle 
  e^{[ \mathcal C^v(t)- \mathcal C^c(t)]\otimes\mathbb I +
  \mathbb I\otimes[ \mathcal C^c(0)- \mathcal C^v(0)]}
  \rangle_{kk';qq'},
\end{align} 
where $kk'qq'$ matrix element of the tensor product $A\otimes B$ is given by $A_{kk'}B_{qq'}$. We also introduce the following notations for later convenience,
\begin{align}
    \mathcal C^{cv}(t)&=  \mathcal C^c(t)- \mathcal C^v(t)=d_{cv}^\dagger(t)-d_{cv}^T(t)\label{eq:dcv}\\
  (d_{cv}^\dagger)_{kk'}(t)&=\sum_\alpha g^{cv}_{\vec{k-k'},\alpha}D^\dagger_{\vec{k-k'},\alpha}e^{i\Omega_\alpha t}\\
  (d_{cv}^T)_{kk'}(t)&=\sum_\alpha g^{cv}_{\vec{k-k'},\alpha}D_{\vec{k'-k},\alpha}e^{-i\Omega_\alpha t}\\
  g^{cv}_{\vec{k-k'},\alpha}&=g^{c}_{\vec{k-k'},\alpha}-g^{v}_{\vec{k-k'},\alpha}.
\end{align}
Inserting Eq.(\ref{eq:dcv}) into Eq.(\ref{eq:thavg_ph}), we obtain
\begin{align}
  &\left\langle 
  (e^{- \mathcal C^{cv}(t)})_{kk'} (e^{ \mathcal C^{cv}(0)})_{qq'}
  \right\rangle\nonumber\\
  &\quad=\left\langle 
  e^{-[d_{cv}^\dagger(t)\otimes\mathbb I-\mathbb I\otimes d_{cv}^\dagger(0)] +
  [d_{cv}^T(t)\otimes\mathbb I-\mathbb I\otimes d_{cv}^T(0)]}
  \right\rangle_{kk';qq'}.
\end{align} 
The first term in the exponential only contains the creation operators while the second term only contains the annihilation operators; they are hermitian conjugates with commutator being central. The cumulant expansion of the following type is applied,
\begin{align}
  \left\langle \exp(-\mathcal O^\dagger+\mathcal O) \right\rangle
  = e^{-\frac{1}{2}\langle \mathcal O^\dagger\mathcal O +\mathcal O \mathcal O^\dagger\rangle},
\end{align}
where the expansion is truncated at the quadratic order as the density matrix of phonons $\rho=\frac{1}{Z}e^{-\beta\bar H_{ph}^0}$ is quadratic in the phonon operators. Using the matrix product identity $(A\otimes B)(C\otimes D)=(AC)\otimes (BD)$, we rewrite Eq.(\ref{eq:thavg_ph}) as 
\begin{align}
    &\left\langle 
  (e^{- \mathcal C^{cv}(t)})_{kk'} (e^{ \mathcal C^{cv}(0)})_{qq'}
  \right\rangle\nonumber\\
  = &[e^{\langle d_{cv}^T(t)\otimes d_{cv}^\dagger(0)
  +d_{cv}^\dagger(t)\otimes d_{cv}^T(0)\rangle}]_{kk';qq'}[e^{-\frac{1}{2}\langle \{d_{cv}^\dagger(t),d_{cv}^T(t)\}\rangle}]_{kk}
  [e^{-\frac{1}{2}\langle \{d_{cv}^\dagger(0),d_{cv}^T(0)\}\rangle}]_{q'q'}
\end{align}
where $\{\cdot\ ,\cdot\}$ denotes the anti-commutator and we have used the fact that $e^{-\frac{1}{2}\langle\{d_{cv}^\dagger(t),d_{cv}^T(t)\}\rangle}$
is diagonal in $kk'$ indices and $e^{-\frac{1}{2}\langle \{d_{cv}^\dagger(0),d_{cv}^T(0)\}\rangle}$ is diagonal in $qq'$. The diagonal terms are evaluated to yield
\begin{align}\label{eq:thavg_ph13}
  &[e^{-\frac{1}{2}\langle \{d_{cv}^\dagger(t),d_{cv}^T(t)\}\rangle}]_{kk}
  =[e^{-\frac{1}{2}\langle \{d_{cv}^\dagger(0),d_{cv}^T(0)\}\rangle}]_{q'q'}\nonumber\\
  &=e^{-\frac{1}{2}\sum_{\vec{q}\alpha}[2n_{B}(\Omega_\alpha,T)+1]|g^{cv}_{\vec q\alpha}|^2}
  \equiv e^{-\frac{1}{2}\tilde{G}_{cv}(T)}.
\end{align}
Here we have defined $\tilde{G}_{cv}(T)=\sum_{\vec{q}\alpha}[2n_{B}(\Omega_\alpha,T)+1]|g^{cv}_{\vec q\alpha}|^2$, which resembles the Huang-Rhys factor at finite temperatures. Following Ref. [\onlinecite{munn1985theory}], one can calculate order by order to obtain 
\begin{align}\label{eq:thavg_ph2}
  &[e^{\langle d_{cv}^T(t)\otimes d_{cv}^\dagger(0)
  +d_{cv}^\dagger(t)\otimes d_{cv}^T(0)\rangle}]_{kk';qq'}
  =\delta_{k-k',q'-q}\left(e^{\sum_{\alpha,\pm} \pm G_{cv}^\alpha n_{B}(\pm\Omega_\alpha,T)\exp(\pm i\Omega_\alpha t)}
   \right)_{kk'} 
\end{align}
where the matrix element of $G_{cv}^\alpha$ is given by $G_{cv,\vec k\vec k'}^\alpha
=|g^c_{\vec k-\vec k',\alpha}-g^v_{\vec k-\vec k',\alpha}|^2$. Combining Eq.(\ref{eq:thavg_ph13})
with Eq.(\ref{eq:thavg_ph2}), and swapping the indices using $G_{cv,\vec k\vec k'}^\alpha=G_{cv,\vec k'\vec k}^\alpha$, we have 
\begin{align}
  &\langle 
  (e^{ \mathcal C^v(t)- \mathcal C^c(t)})_{\vec {k_1k_2}} (e^{ \mathcal C^c(0)-  \mathcal C^v(0)})_{\vec {k_4k_3}}
  \rangle= e^{-\tilde{G}_{cv}(T)}(e^{G_{cv}(T,t)})_{\vec {k_2k_1}}\delta_{\vec {k_1-k_2},\vec {k_4-k_3}}
\end{align}
where the matrix $G_{cv}(T,t)$ is defined through
\begin{align}
  [G_{cv}(T,t)]_{\vec k\vec k'}= \sum_\alpha G_{cv,\vec k\vec k'}^\alpha[ n_{B}(\Omega_\alpha,T)e^{i\Omega_\alpha t}
  +(n_{B}(\Omega_\alpha,T)+1)e^{-i\Omega_\alpha t}].
\end{align}
Note that this thermal average of phonon operators reduces to the previous zero-temperature result in the limit $n_B(\Omega_\alpha,T\rightarrow0)\rightarrow 0$.

For the thermal average of four-polaron operators, we have the following relation:
\begin{align}
  a^\dagger_{\vec k_1 v}(t) a_{\vec k_2 c}(t)
  =e^{it\cdot\textup{ad}_{\bar H_p}}a^\dagger_{\vec k_1 v}a_{\vec k_2 c}.
\end{align}
The adjoint action of $\bar H_p$ defined through the commutator $[\bar H_p, \ \cdot \ ]$ on the polaron bilinear operators in general leads to hierarchy problems in which truncation should be taken at certain order. To avoid this challenge, we follow Ref.~[\onlinecite{hannewald2005nonperturbative}] and perform mean-field approximation, yielding
\begin{align}
 \textup{ad}_{\bar H_p} a^\dagger_{\vec k_1 v}a_{\vec k_2 c}
  &\approx -\sum_{\vec k'_1\vec k'_2} 
  K_{\vec k_1 \vec k_2;\vec k'_1\vec k'_2}a^\dagger_{\vec k'_1 v}a_{\vec k'_2 c}
\end{align}
where $ K_{\vec k_1 \vec k_2;\vec k'_1\vec k'_2}=[(\tilde e_{\vec k_2c}-\tilde e_{\vec k_1v})\delta_{\vec k_1 \vec k_1'}+(n_{\vec k_2c}-n_{\vec k_1v})V^{vc}_{\vec k_1-\vec k_1'}]\delta_{\vec k_1-\vec k'_1,\vec k_2-\vec k'_2}$, and we have ignored the Hartree term containing $V_{\vec q=0}$ whose contribution to polaron dispersion is just a constant shift (and it cancels out when $V^{cv}=V^{vc}, V^{cc}=V^{vv}$). Here $K$ is the kernel, $\tilde e_{\vec k\lambda}=e_{\vec k\lambda}-\sum_{\vec{k'}}V^{\lambda\lambda}_{\vec{k-k'}}n_{\vec{k'}\lambda}$ is the polaron dispersion with Fock term correction, $n_{\vec k\lambda}$ is the fermionic occupation number. For materials with band gaps larger than 1eV, we can further make the ``large-band-gap approximation" for the fermionic occupation number, i.e.
\begin{align}
  n_{\vec kc}\approx 0\ \forall c, \qquad
  n_{\vec kv}\approx 1\ \forall v,
\end{align}
which remains valid for temperatures up to $10^4$K \cite{antonius2022theory}. To simplify the notion, we treat $(\vec k_1 \vec k_2;\vec k'_1\vec k'_2)$ as the entry of the matrix element 
$K_{\vec k_1 \vec k_2;\vec k'_1\vec k'_2}$ giving
\begin{align}
  a^\dagger_{\vec k_1 v}(t) a_{\vec k_2 c}(t)
  \approx
  \sum_{\vec k'_1\vec k'_2} 
  (e^{-it\cdot K})_{\vec k_1 \vec k_2;\vec k'_1\vec k'_2}
  a^\dagger_{\vec k'_1 v}a_{\vec k'_2 c}.
\end{align}

We combine the thermal average of four-polaron operators and that of the exponential of phononic operators assuming  $\langle a^\dagger_{\vec k'_1 v}a_{\vec k'_2 c} a^\dagger_{\vec k_3 c} a_{\vec k_4 v}\rangle= n_{\vec k_4v}(1-n_{\vec k_3c}) \delta_{\vec k_3,\vec k'_2}\delta_{\vec k_4,\vec k'_1}\approx \delta_{\vec k_3,\vec k'_2}\delta_{\vec k_4,\vec k'_1} $, which leads to 
\begin{align}\label{eq:P_model1}
  P^R_>(t)
  &=-i\frac{\Theta(t)}{N}e^{-\tilde G_{cv}(T)} \sum_{\vec {k_1...k_4}}
  (e^{-it\cdot K})_{\vec k_1 \vec k_2;\vec k_4\vec k_3}
  (e^{G_{cv}(T,t)})_{\vec k_1,\vec k_2}\nonumber\\
  &=-i\frac{\Theta(t)}{N}
  \langle e^{G_{cv}(T,t)}|e^{-it\cdot K}|e^{-\tilde G_{cv}(T)}\rangle.
\end{align}
Here we explain the notation: the inner product of two bilocal functions in momentum space $f$, $g$  is given by $\langle f|g\rangle=\sum_{k_1k_2}f_{k_1,k_2}g_{k_1,k_2}$, and the action of a linear functional $A$ is $A|f\rangle_{k_1,k_2}=\sum_{k_3,k_4}A_{k_1k_2;k_3k_4}f_{k_3,k_4}$, and $|e^{-\tilde G_{cv}}\rangle$ is a constant function with value $e^{-\tilde G_{cv}}$ in momentum space. As a byproduct, we identify the eigenvalue equation of the kernel $K$ as the Wannier equation, i.e.
\begin{align}
  K|\psi_{\mu}\rangle=E_\mu|\psi_{\mu}\rangle,
\end{align}
where $\mu$ is a collective index including continuous center-of-mass momentum indices labeled by $\vec Q$ and a set of discrete quantum numbers $\nu$, i.e. $\mu=(\vec Q,\nu)$. The eigenstate is a tensor product of c.o.m wavefunction and relative-motion wave function $|\psi_{\vec Q\nu}\rangle=|\vec Q\rangle\otimes|\psi_\nu\rangle$.

\subsubsection{Absorption spectrum for single phonon mode and constant EPC}
For analytic calculation, we consider a special case where only one optical phonon mode contributes to the absorption spectrum. We take EPC to be $g^\lambda_\vec{kk'}=g_\lambda/\sqrt{N}$, where $N$ is the number of unit cells. Then the $G_{cv}(T,t)$ is a $N\times N$ matrix with all elements being 
\begin{align}
  G_{cv,\vec{kk'}}(T,t)=\frac{1}{N}g_{cv}^2\left[n_B(\Omega,T)e^{i\Omega t}+(n_B(\Omega,T)+1)e^{-i\Omega t}\right],
\end{align}
where $g_{cv}=g_c-g_v$. The exponential of $G_{cv}(T,t)$ is given by 
\begin{align}
  (e^{G_{cv}(T,t)})_{\vec k\vec k'}
  =\delta_{\vec k,\vec k'}+\frac{1}{N}\left(e^{g_{cv}^2\left[n_B(\Omega,T)
  e^{i\Omega t}+(n_B(\Omega,T)+1)e^{-i\Omega t}\right]}-1\right),
\end{align}
Similarly, we have
\begin{align}
  (e^{-\tilde G_{cv}(T,t)})_{\vec k,\vec k'}
  =e^{-g_{cv}^2(2n_B(\Omega,T)+1)}.
\end{align}
Inserting the resolution of identity $\mathbb I=\sum_{\vec Q\nu }\ket{\vec Q\nu}\bra{\vec Q\nu}$ into the absorption coefficient Eq.(\ref{eq:P_model1}), and applying the relation $\langle\vec k\vec k'|\vec Q\nu \rangle=\delta_{\vec Q,\vec k-\vec k'}\psi_\nu(\frac{m_h}{M}\vec k+\frac{m_e}{M}\vec k')$ ($m_p, m_a$ is the mass of polaron and anti-polaron, $M$ is the mass of polaron pair), along with $\psi_\nu(\vec r=\vec 0)=\frac{1}{N}\sum_{\vec k}\psi_\nu(\vec k)$, 
\begin{align}
  P^R_>(t)
  &=-i\Theta(t)\sum_{\vec Q\nu} |\psi_\nu(\vec r=\vec 0)|^2
  e^{-g_{cv}^2(2n_B(\Omega,T)+1)}\left(\sum_{\vec Q}\underbrace{(N\delta_{\vec Q,0}-1)e^{-iE_{\vec Q\nu}t}}_{\approx0}\right.\nonumber\\
  &\qquad\left.+ \sum_{\vec Q} e^{g_{cv}^2\left[n_B(\Omega,T)e^{i\Omega t}+(n_B(\Omega,T)+1)e^{-i\Omega t}\right]-iE_{\vec Q\nu}t}\right),
\end{align}
where the first term in the parentheses is approximately zero under the flatband approximation $E_{\vec Q\nu}\approx E_{\vec 0\nu}$, and $|\psi_\nu(\vec r=\vec 0)|^2$ is probability that polaron and anti-polaron spatially overlap. We then expand time dependent exponential $e^{g_{cv}^2\left[n_B(\Omega,T)e^{i\Omega t}+(n_B(\Omega,T)+1)e^{-i\Omega t}\right]}$  using the generating function of modified Bessel function $e^{\frac{1}{2}z(t+t^{-1})}=\sum_{m\in\mathbb Z}t^mI_m(z)$.  Defining $t=e^{\frac{1}{2}\beta\Omega-it\Omega}$, and $z=\frac{g_{cv}^2}{2\sinh(\beta\Omega/2)}=\sqrt{n_B(\Omega,T)(n_B(\Omega,T)+1)}g_{cv}^2$, we obtain
\begin{align}
  e^{g_{cv}^2\left[n_B(\Omega,T)e^{i\Omega t}+(n_B(\Omega,T)+1)e^{-i\Omega t}\right]}
  =\sum_{m\in\mathbb Z}e^{\frac{1}{2}m\beta\Omega-im\Omega t}
  I_m\left(\frac{g_{cv}^2}{\sinh(\beta\Omega/2)}\right)
\end{align}
Therefore, the Fourier transformation $P^R_>(\omega)=\int_t P^R(t)e^{i\omega^+ t}$ is given by
\begin{align}
    P^R_>(\omega,T)
    =e^{-g_{cv}^2(2n_B(\Omega,T)+1)}
    \sum_{\vec Q\nu} \sum_{m=-\infty}^\infty
    \frac{e^{\frac{1}{2}m\beta\Omega}|\psi_\nu(\vec r=0)|^2}{\omega^+-m\Omega-E_{\vec Q\nu}}
    I_m\left(\frac{g_{cv}^2}{\sinh(\beta\Omega/2)}\right).
\end{align}
 We now obtain the absorption spectrum at temperature $T$
\begin{align}
    \label{eq:LR_Absorption}
    \alpha(\omega)
    =e^{-g_{cv}^2(2n_B(\Omega,T)+1)}
    \sum_{\vec Q\nu} \sum_{m=-\infty}^\infty
    \frac{e^{\frac{1}{2}m\beta\Omega}|\psi_\nu(\vec r=0)|^2\gamma_m}{(\omega-m\Omega-E_{\vec Q\nu})^2+\gamma_m^2}
    I_m\left(\frac{g_{cv}^2}{\sinh(\beta\Omega/2)}\right)
\end{align}
where we have added an $m$-dependent imaginary part $i\gamma_m$ to account for higher-order scattering-induced dephasing.
The zero temperature limit can be obtained by noting that $n_B(\Omega,T\rightarrow 0)=0$, $I_m\left(\frac{g_{cv}^2}{\sin (\beta\Omega/2)}\right)\sim\frac{1}{\Gamma(m+1)}\left(\frac{g_{cv}^2}{2\sinh(\beta\Omega/2)}\right)^m$,
\begin{align}
    P^R_>(\omega,T=0)
    &=e^{-g_{cv}^2}
    \sum_{\vec Q\nu} \sum_{m=0}^\infty \frac{g_{cv}^{2m}}{m!}
    \frac{|\psi_\nu(\vec r=0)|^2}{\omega^+-m\Omega-E_{\vec Q\nu}},
\end{align}
where only $m>0$ terms survive since $1/\Gamma(z+1)=0$ at negative integers. 

\subsection{Equivalence of the two approaches}
To show that Eq.~(\ref{eq:EOM_Amplitude}) gives the correct side band amplitude, we return to Eq.~(\ref{eq:Corr_EOM}). In general, it has the form:
\begin{align}
    [f_{c,v}(\omega)-E^{c,v}_{\{\alpha,\beta\}}]\bar{\Pi}^{c,v}_{\{\alpha,\beta\}}(\vk,\vq) &= S_{\{\alpha,\beta\}}(\vk,\vq)
\end{align}
where $E^{c,v}_{\{\alpha,\beta\}}$ is the net phonon absorption/emission energy of a particular configuration of absorbed and emitted phonons, and for convenience we have simplified the remaining terms. Applying the commutator factor $C^{\alpha,\beta}_{c,v}$ and summing over all configurations, we have:
\begin{align}
    \sum_{\alpha,\beta}C^{\alpha,\beta}_{c,v}\sum_{\{\alpha,\beta\}}[f_{c,v}(\omega)-E^{c,v}_{\{\alpha,\beta\}}]\bar{\Pi}^{c,v}_{\{\alpha,\beta\}}(\vk,\vq) &= A^{v,c}_{\vk,\vq}\expval{(e^{C_v-C_c})_{-\vq,0}(e^{C_c-C_v})_{\vq,0}}\label{eq:epiphany}
\end{align}
where:
\begin{align}
    A^{v,c}_{\vk,\vq} &\equiv \mathcal{F}^{v,c}\Big\{f^c_{\vk-\vq}(1-f^v_\vk) + \expval{B^\dagger B}(f^c_{\vk-\vq}-f^v_\vk)\Big\}
\end{align}
The R.H.S. of Eq.(\ref{eq:epiphany}) comes from the original definition of $S_{\{\alpha,\beta\}}(\vk,\vq)$ and $C^{\alpha,\beta}_{c,v}$, and can be seen as the result of calculating contributions to the EOM from $\bar{H}_{pol-em}$ without splitting up the phonon operators into absorption and emission pieces. We recognize the phonon operator on the R.H.S. of Eq.(\ref{eq:epiphany}) as the one evaluated in Eq.(\ref{eq:thavg_ph}), yielding:
\begin{align}
    &e^{-\bar{G}_{c,v}-\tfrac{1}{2}\tilde{G}_{c,v}}\bigg(e^{\bar{G}_{c,v}+\mathbb{1}}e^{\bar{G}_{c,v}}\bigg)_{\vq,0} = \sum_{\alpha,\beta}\frac{1}{\alpha !\beta !}[(\bar{G}_{c,v}+\mathbb{1})^\alpha(\bar{G}_{c,v})^\beta]_{\vq,0}\nonumber\\
    &= e^{-\bar{G}_{c,v}-\tfrac{1}{2}\tilde{G}_{c,v}}\sum_{\alpha,\beta}\frac{1}{\alpha !\beta !}\sum_{\{\alpha,\beta\}}\prod_{i=0}^{\alpha-1}(\bar{G}_{c,v}+\mathbb{1})_{\nu_i,\vq'_i}\prod_{j=\alpha}^{\alpha+\beta-1}(\bar{G}_{c,v})_{\nu_l,\vq'_j}
\end{align}
The summations on both sides of Eq.(\ref{eq:epiphany}) are of the same form, and matching terms element wise yields the side-band amplitude given by Eq.(\ref{eq:EOM_Amplitude}). For momentum independent phonon dispersion this  yields the absorption spectrum given by Eq.(\ref{eq:LR_Absorption}). 


\section{Model II: Phonon-assisted intraband and interband absorption}

\subsection{Derivation of the transformed Hamiltonian}\label{ap_mod2}
The Hamiltonian in this case reads $H=H^0_e+H^0_{ph}+H^0_{em}+H_{e-e}+H_{e-ph}+H_{e-em}$ with
\begin{align}
  H^0&=\sum_{\vec k\lambda} \epsilon_{\vec k \lambda} a_{\vec k\lambda}^\dagger a_{\vec k\lambda}
  +\sum_{\vec q\alpha} \Omega_{\vec q\alpha} D_{\vec q\alpha}^\dagger D_{\vec q\alpha}
  +\sum_{\vec q} \omega_{\vec q } B_{\vec q}^\dagger B_{\vec q}\\
  H_{e-e}&=\frac{1}{2}\sum_{\vec {kk'q}\lambda\lambda'}V_{\vec q}a^\dagger_{\vec{k-q}\lambda}
  a^\dagger_{\vec{k'+q}\lambda'}a_{\vec k'\lambda'}a_{\vec k\lambda}\\
  H_{e-ph}&=\sum_{\vec {kq} \lambda \lambda'\alpha}\Omega_{\vec q\alpha}g^{\lambda\lambda'}_{\vec q\alpha}
  a^\dagger_{\vec {k-q},\lambda'}a_{\vec {k}\lambda}(D_{-\vec q,\alpha}+D^\dagger_{\vec q\alpha})\\
  H_{e-em}&=-\sum_{\vec {kq}\lambda}
           i\mathcal{F}^{\lambda\bar\lambda}_{\vec q} 
           a^\dagger_{\vec {k+q}\lambda}a_{\vec {k}\bar\lambda}B_{\vec q}+h.c..
\end{align}
The unitary transformation parallels the previous case, and is given by 
\begin{align}
  U
  =\exp[\sum_{\vec{kq}\lambda\lambda'\alpha}g^{\lambda\lambda'}_{\vec q \alpha}Q_{\vec q\alpha}
  a_{\vec {k-q,\lambda'}}^\dagger a_{\vec k\lambda}],
\end{align}

We now derive the renormalized Hamiltonian. For later convenience we first derive the adjoin action of $S$ on the creation/annihilation operators of electrons, phonons and photons.
\begin{itemize}
  \item  The action of $\textup{ad}_S$ on $a_{\vec k\lambda}$ is given by 
  \begin{align}
    \textup{ad}_S a_{\vec k\lambda}
    &=-\sum_{\vec {p}\alpha\lambda'}g^{\lambda'\lambda}_{\vec{p-k},\alpha}Q_{\vec{p-k},\alpha}a_{\vec p\lambda'}
    =-\sum_{\vec p\lambda'}a_{\vec p\lambda'}\mathcal C_\vec{p\lambda',k\lambda}\\
    \textup{ad}_S a^\dagger_{\vec k\lambda}
    &=\sum_{\vec {p}\alpha \lambda'}g^{\lambda\lambda'}_{\vec{k-p},\alpha}Q_{\vec{k-p},\alpha}a^\dagger_{\vec p\lambda'}
    =\sum_{\vec p\lambda'}\mathcal C_\vec{k\lambda,p\lambda'}a^\dagger_{\vec p\lambda'}
  \end{align}
  where we have defined $\mathcal C_\vec{k\lambda,k'\lambda'}\equiv \sum_\alpha g^{\lambda\lambda'}_{\vec{k-k'},\alpha} Q_{\vec{k-k'},\alpha}$ as the matrix element of the anti-Hermitian operator $\mathcal C$. 
  This allows to write the write the adjoin action of $U=e^S$ on electronic operators as
  \begin{align}
    \textup{Ad}_Ua_{\vec k\lambda}
    &=e^{\textup{ad}_S} a_{\vec k\lambda}
    =\sum_{\vec p\lambda'}a_{\vec p\lambda'}(e^{-\mathcal{C}})_{\vec{p\lambda',k\lambda}}\label{eq:a}\\
    \textup{Ad}_Ua^\dagger_{\vec k\lambda}
    &=e^{\textup{ad}_S} a^\dagger_{\vec k\lambda}
    =\sum_{\vec p\lambda'}(e^{\mathcal{C}})_{\vec{k\lambda,p\lambda'}}a^\dagger_{\vec p\lambda'}\label{eq:a_dagger}
  \end{align}
  \item  The action of $\textup{ad}_S$ on $D_{\vec q\alpha}$ is written as
  \begin{align}
    \textup{ad}_S D_{\vec q\alpha} &=-\sum_{\vec{k}\lambda\lambda'}
     g^{\lambda\lambda'}_{\vec q\alpha} a_{\vec {k-q,\lambda'}}^\dagger a_{\vec k\lambda}\\
     \textup{ad}_S D^\dagger_{\vec q\alpha} &=-\sum_{\vec{k}\lambda\lambda'}
     g^{\lambda\lambda'}_{-\vec q\alpha} a_{\vec{k+q} \lambda'}^\dagger a_{\vec{k}\lambda},
  \end{align}
  which yields $\textup{ad}_S \sum_{\vec q\alpha}\Omega_{\vec q\alpha}D^\dagger_{\vec q\alpha}D_{\vec q\alpha} 
  =-\sum_{\vec{k}\vec q\alpha\lambda\lambda'}
   g^{\lambda\lambda'}_{\vec q\alpha} \Omega_{\vec q\alpha}
   a_{\vec {k-q,\lambda'}}^\dagger a_{\vec k\lambda}
   (D^\dagger_{\vec q \alpha}+D_{-\vec q,\alpha})$ assuming $\Omega_{\vec q\alpha}=\Omega_{-\vec q\alpha}$, and thus $\textup{ad}_SH^0_{ph}+H_{e-ph}=0$ as promised.
  \item  Since $S$ commutes with $B_{\vec q}$, then
  \begin{align}
    \textup{ad}_S B_{\vec q} =0,\qquad e^{\textup{ad}_S} B_{\vec q} =0.
  \end{align}
\end{itemize}
We are ready to give the expression for the transformed Hamiltonian:
\begin{itemize}
  \item[(1)] To make the notation compact, we define matrix element 
  $\mathcal{E}_{\vec {k\lambda,k'\lambda'}}=\epsilon_{\vec k\lambda}\delta_{\vec {kk'}}\delta_{\lambda\lambda'}$.
  Then, Eq.(\ref{eq:a}) and Eq.(\ref{eq:a_dagger}) lead to 
  \begin{align} 
    e^{\textup{ad}_S} H^0_e
    &=\sum_{\vec k\lambda} \epsilon_{\vec k \lambda} 
     (e^{\textup{ad}_S}a_{\vec k\lambda}^\dagger) (e^{\textup{ad}_S}a_{\vec k\lambda})\nonumber\\
    &=\sum_{\vec {pp'k}\lambda\lambda'\lambda''} \epsilon_{\vec k \lambda} 
    (e^{\mathcal{C}})_{\vec{k\lambda,p\lambda'}}a^\dagger_{\vec p\lambda'}
    a_{\vec p'\lambda''}(e^{-\mathcal{C}})_{\vec{p'\lambda'',k\lambda}}\nonumber\\
    &=\sum_{\vec {pp'k}\lambda\lambda'}(e^{-\mathcal{C}}\mathcal{E} e^{\mathcal{C}})_{\vec{p'}\lambda',\vec{p}\lambda}
    a^\dagger_{\vec p\lambda}a_{\vec p'\lambda'}.
  \end{align}
  \item[(2)]  We then calculate $\textup{ad}_S H_{e-ph}$, which is given by
  \begin{align}
    \textup{ad}_S H_{e-ph}
    &=\textup{ad}_S \sum_{\vec {kq} \lambda\lambda'\alpha}\Omega_{\vec q\alpha}g^{\lambda\lambda'}_{\vec q\alpha}
    a^\dagger_{\vec {k-q},\lambda'}a_{\vec {k}\lambda}(D_{-\vec q,\alpha}+D^\dagger_{\vec q\alpha})
    =(\textup{ad}_S H_{e-ph})^{(1)}+(\textup{ad}_S H_{e-ph})^{(2)}
  \end{align}
  \begin{align}
    (\textup{ad}_S H_{e-ph})^{(1)}
    &=-2\sum_{\vec{kk'q}\{\lambda_i\}\alpha} \Omega_{\vec q\alpha}
    g^{\lambda_1\lambda_2}_{\vec q\alpha}(g^{\lambda_4\lambda_3}_{\vec q\alpha})^*
    a^\dagger_{\vec{k-q}\lambda_2}a_{\vec k\lambda_1}
    a^\dagger_{\vec{k'+q}\lambda_4}a_{\vec k' \lambda_3}\nonumber \\
    &=-2\sum_{\vec{kq}\lambda\lambda'\tilde\lambda\alpha} \Omega_{\vec q\alpha}g^{\tilde\lambda\lambda'}_{\vec q\alpha}(g^{\tilde\lambda\lambda}_{\vec q\alpha})^*
    a^\dagger_{\vec{k}\lambda'}a_{\vec {k}\lambda}
    \nonumber
    \\
    &\qquad-2\sum_{\vec{kk'q}\{\lambda_i\}\alpha} \Omega_{\vec q\alpha}
    g^{\lambda_1\lambda_2}_{\vec q\alpha}(g^{\lambda_4\lambda_3}_{\vec q\alpha})^*
    a^\dagger_{\vec{k-q}\lambda_2}a^\dagger_{\vec{k'+q}\lambda_4}
    a_{\vec k' \lambda_3}a_{\vec k\lambda_1}.
  \end{align}
  \begin{align}
    (\textup{ad}_S H_{e-ph})^{(2)}
    &=\sum_{\vec k\vec q\vec q'\alpha\alpha'\{\lambda_i\}}
    \left(
      g^{\lambda_1\lambda_2}_{\vec q\alpha} g^{\lambda_3\lambda_1}_{\vec q'\alpha'}
      -
      g^{\lambda_1\lambda_2}_{\vec q'\alpha'} g^{\lambda_3\lambda_1}_{\vec q\alpha}
    \right)
    \delta_{\lambda_1,\lambda_4}
    a^\dagger_{\vec{k}-\vec{q},\lambda_2}a_{\vec{k}+\vec{q}',\lambda_3}
    \Omega_{\vec q'\alpha'}\nonumber\\
    &\qquad\qquad\qquad
    (D^\dagger_{\vec q\alpha}-D_{-\vec q,\alpha})(D_{-\vec q',\alpha'}+D^\dagger_{\vec q'\alpha'}).
  \end{align}
  When $g^{\lambda\lambda'}$ is diagonal in the band indices, $(\textup{ad}_S H_{e-ph})^{(2)}$ vanishes identically. More generally, however, this term remains finite. For the case of momentum independent coupling relevant here $(\textup{ad}_S H_{e-ph})^{(2)}$ vanishes identically. This in turn leads to a simplification analogous to model I namely,

 
  \begin{align}
    (\textup{ad}_S)^2 H_{e-ph}= 0 \qquad (\textup{ad}_S)^n H_{e-ph}=0, \textup{ for }\forall n\ge2,
  \end{align} 
  which can be derived by rewriting $S$ and $\textup{ad}_S H_{e-ph}$ in real space. Notice that 
  \begin{align*}
    \textup{ad}_S H_{e-ph}
    &=-2\sum_{ \{\lambda_i\} \alpha}\int_{\vec {r,r'}}
    [\Omega_\alpha\star g^{\lambda_1\lambda_2}_{\alpha}\star (g^{\lambda_4\lambda_3}_{\alpha})^*](\vec r-\vec r')
    P_{\lambda_2\lambda_1 }({\vec r})P_{\lambda_4\lambda_3}({\vec r'})\\
    S&=\sum_{\lambda\lambda'\alpha}\int_{\vec r}
    [g^{\lambda\lambda'}_{\alpha}\star Q_\alpha](\vec r)P_{\lambda'\lambda}(\vec r),
  \end{align*}
  where $[\Omega_\alpha\star g^{\lambda_1\lambda_2}_{\alpha}\star (g^{\lambda_4\lambda_3}_{\alpha})]$ and $[g^{\lambda\lambda'}_{\alpha}\star Q_\alpha]$ are the convolutions of the real space representation of each term inside, and $P_{\lambda\lambda'}({\vec r})=c^\dagger_{\lambda}(\vec r)c_{\lambda'}(\vec r)$. Then the commutation relation of density operator $[P_{\lambda\lambda'}({\vec r}),P_{\lambda_1\lambda_2}({\vec r'})]
  =[ P_{\lambda,\lambda_2}(\vec r)\delta_{\lambda'\lambda_1}
  -P_{\lambda_1,\lambda'}(\vec r)\delta_{\lambda\lambda_2}]\delta(\vec r-\vec r')$ leads to 
  \begin{align*}
    \textup{ad}_S^2 H_{e-ph}
    &=-2\sum_{ \{\lambda\} \alpha\alpha'}\int_{\vec {r,r',r''}}
    [\Omega_\alpha\star g^{\lambda_1\lambda_2}_{\alpha}\star (g^{\lambda_4\lambda_3}_{\alpha})^*](\vec r-\vec r')
    [g^{\lambda\lambda'}_{\alpha'}\star Q_{\alpha'}](\vec r'')
    \\
    &\qquad\qquad
    [P_{\lambda'\lambda}(\vec r''), 
    P_{\lambda_2\lambda_1 }({\vec r})
    P_{\lambda_4\lambda_3}({\vec r'})],\\
    &=-2\sum_{ \{\lambda\} \alpha\alpha'}\int_{\vec {r,r',r''}}
    [\Omega_\alpha\star g^{\lambda_1\lambda_2}_{\alpha}\star (g^{\lambda_4\lambda_3}_{\alpha})^*](\vec r-\vec r')
    [g^{\lambda\lambda'}_{\alpha'}\star Q_{\alpha'}](\vec r'')
    \\
    &\qquad\qquad
      \left\{
      [ P_{\lambda',\lambda_1}(\vec r)\delta_{\lambda\lambda_2}
      -P_{\lambda_2,\lambda}(\vec r)\delta_{\lambda'\lambda_1}]P_{\lambda_4\lambda_3}({\vec r'})\delta(\vec r-\vec r'')
    \right.
    \\
    &\qquad\qquad
    \left.
    +P_{\lambda_2\lambda_1 }({\vec r})
    [
    P_{\lambda',\lambda_3}(\vec r')\delta_{\lambda\lambda_4}
    -P_{\lambda,\lambda_4}(\vec r')\delta_{\lambda'\lambda_3}
    ]
    \delta(\vec r''-\vec r')
    \right\}\\
    &=
    (\textup{ad}_S^2 H_{e-ph})^{(1)}
    +
    (\textup{ad}_S^2 H_{e-ph})^{(2)}
  \end{align*}
  Based on the assumption, electron-phonon coupling is symmetric for band indices,
  namely $g^{\lambda\lambda'}_{\alpha}(\vec r) = g^{\lambda'\lambda}_{\alpha}(\vec r)$, this is equivalent to the condition that $g^{\lambda\lambda'}_{\alpha}(\vec r)$ is real in the real space representation as $g^{\lambda\lambda'}_{\alpha}(\vec r)\in \mathbb R \rightarrow g^{\lambda\lambda'}_{\vec q \alpha}=(g^{\lambda\lambda'}_{-\vec q \alpha})^*$.  Using this symmetry condition, we see the exact cancellation between polarization operators $P$ if we further assume $g^{\lambda\lambda'}_{\vec q\alpha}$ is flat in the momentum space. For example the first term $(\textup{ad}_S^2 H_{e-ph})^{(1)}$
  \begin{align*}
    (\textup{ad}_S^2 H_{e-ph})^{(1)}
    &=-2\sum_{ \{\lambda\} \alpha\alpha'}\int_{\vec {r}}
    \Omega_\alpha g^{\lambda_1\lambda_2}_{\alpha}(g^{\lambda_4\lambda_3}_{\alpha})^* g^{\lambda\lambda_1}_{\alpha'}
       Q_{\alpha'}(\vec r)
       [ P_{\lambda,\lambda_2}(\vec r)-P_{\lambda_2,\lambda}(\vec r)]
      P_{\lambda_4\lambda_3}({\vec r})
  \end{align*}
  Since $P_{\lambda,\lambda_2}(\vec r)-P_{\lambda_2,\lambda}(\vec r)$ is anti-symmetric, this term vanishes identically as long as $(g_{\alpha'}g_{\alpha})^{\lambda\lambda_2}=\sum_{\lambda_1}g^{\lambda\lambda_1}_{\alpha'}g^{\lambda_1\lambda_2}_{\alpha}$ is symmetric with respect to $\lambda\leftrightarrow\lambda_2$. Indeed, $\sum_{\lambda_1}g^{\lambda\lambda_1}_{\alpha'}g^{\lambda_1\lambda_2}_{\alpha}=\sum_{\lambda_1}g^{\lambda_1\lambda}_{\alpha'}g^{\lambda_2\lambda_1}_{\alpha}=(g_{\alpha}g_{\alpha'})^{\lambda_2\lambda}$. Therefore, if the coupling matrix commute for different phonon branches or if we only consider one single phonon branch, then we have $(g_{\alpha'}g_{\alpha})^{\lambda\lambda_2}=(g_{\alpha'}g_{\alpha})^{\lambda_2\lambda}$ which makes the term $(\textup{ad}_S^2 H_{e-ph})^{(1)}=0$ and similarly $(\textup{ad}_S^2 H_{e-ph})^{(2)}=0$. All higher order adjoint action gives zero.
  \item[(3)] The renormalization of $H^0_{ph}$ and $H_{e-ph}$:
  \begin{align} 
    e^{\textup{ad}_S} H^0_{ph}+e^{\textup{ad}_S} H_{e-ph}
    &=H^0_{ph}+\sum_{n=1}^\infty\frac{1}{n!}(\textup{ad}_S)^n H^0_{ph}
    +\sum_{n=0}^\infty\frac{1}{n!}(\textup{ad}_S)^n H_{e-ph}\nonumber \\
    &=H^0_{ph}+\sum_{n=1}^\infty\frac{n}{(n+1)!}(\textup{ad}_S)^n H_{e-ph}\nonumber\\
    &=H^0_{ph}+\frac{1}{2}\textup{ad}_S H_{e-ph},
  \end{align}
  where we have used the relation $ \textup{ad}_S H^0_{ph}=-H_{e-ph}$, and the fact that 
  $(\textup{ad}_S)^n H_{e-ph}=0, \textup{ for }\forall n\ge2$.
  \item[(4)] The free photon Hamiltonian remains the same: 
  \begin{align} 
    e^{\textup{ad}_S} H^0_{em}=H^0_{em}.
  \end{align}
  \item[(5)] The renormalized photon-electron interaction is written as 
  \begin{align} 
    e^{\textup{ad}_S} H_{e-e}
    &=\frac{1}{2}\sum_{\vec {kk'q}\{\vec k_i\}\lambda\lambda'}V_{\vec q}
    (e^{\mathcal{C}})_{\vec{k-q}\lambda,\vec{k}_1\lambda_1} 
    (e^{\mathcal{C}})_{\vec{k'+q}\lambda',\vec{k}_2\lambda_2} 
    (e^{-\mathcal{C}})_{\vec k_3\lambda_3,\vec k'\lambda'} 
    (e^{-\mathcal{C}})_{\vec k_4\lambda_4,\vec k\lambda} 
    \nonumber \\
    &\qquad \qquad 
    a^\dagger_{{\vec k_1}\lambda_1}
    a^\dagger_{\vec k_2 \lambda_2}
    a_{\vec k_3\lambda_3}
    a_{\vec k_4\lambda_4}.
  \end{align}
Notice that $\sum_{\vec k'\lambda'} (e^{\mathcal{C}})_{\vec{k'+q}\lambda',\vec{k}_2\lambda_2} 
  (e^{-\mathcal{C}})_{\vec k_3\lambda_3,\vec k'\lambda'}=\delta_{\vec{k}_2-\vec q,\vec k_3}\delta_{\lambda_2,\lambda_3}$, which directly follow from the property $\mathcal{C}_{\vec k\lambda,\vec k'\lambda'}=\mathcal{C}_{\vec {k+q}\lambda,\vec {k'+q}\lambda'}$. Therefore, as in Model I, the Coulomb interaction term is not renormalized by $S$, i.e.
  \begin{align}
    e^{\textup{ad}_S} H_{e-e}= H_{e-e}.
  \end{align}
  \item[(6)] We finally calculate the renormalized electron-photon interaction, which is given by 
  \begin{align}
    e^{\textup{ad}_S} H_{e-em}
   &= -\sum_{\vec {kq}\{\vec k_i\}\{\lambda\} }
    i\mathcal{F}^{\lambda\bar\lambda}_{\vec q} 
    (e^{\mathcal{C}})_{\vec{k+q}\lambda,\vec{k}_1\lambda_1} 
    (e^{-\mathcal{C}})_{\vec k_2\lambda_2, \vec k\bar\lambda}
    a^\dagger_{\vec {k}_1\lambda_1}
    a_{\vec {k}_2\lambda_2}B_{\vec q}+h.c. 
  \end{align} 
\end{itemize}
In summary, the transformed total Hamiltonian reads 
\begin{align}
\bar H_{pol}
  &=\sum_{\vec {pp'k}\lambda\lambda'}(e^{-\mathcal{C}}\mathcal{E} e^{\mathcal{C}})_{\vec{p'}\lambda',\vec{p}\lambda}
  a^\dagger_{\vec p\lambda}a_{\vec p'\lambda'}
  -\sum_{\vec{kq}\lambda\lambda'\alpha} \Omega_{\alpha}
  (g_\alpha^2)^{\lambda\lambda'}
  a^\dagger_{\vec{k}\lambda'}a_{\vec {k}\lambda} \\
\bar H^0_{ph}
  &=\sum_{\vec q\alpha}\Omega_{\alpha} D^\dagger_{\vec q\alpha}D_{\vec q\alpha} \qquad\qquad
\bar H^0_{em}
  =\sum_{\vec q}\omega_\vec q B^\dagger_\vec q B_\vec q \\
\bar H_{pol-pol}
  &=\frac{1}{2}\sum_{\vec{kk'q}\lambda\lambda'} 
  (V_\vec{q} \delta_{\lambda_2,\lambda_3}\delta_{\lambda_1,\lambda_4}
  -2\sum_{\alpha}\Omega_{\alpha} 
  g^{\lambda_4\lambda_1}_{\alpha}g^{\lambda_2\lambda_3}_{\alpha})
  a^\dagger_{\vec{k-q}\lambda_1}
  a^\dagger_{\vec{k'+q}\lambda_2}
  a_{\vec k'\lambda_3}
  a_{\vec k\lambda_4}\\
\bar H_{pol-em}
  &=-\sum_{\vec {kq}\{\vec k_i\}\{\lambda\} }
  i\mathcal{F}^{\lambda\bar\lambda}_{\vec q} 
  (e^{\mathcal{C}})_{\vec{k+q}\lambda,\vec{k}_1\lambda_1} 
  (e^{-\mathcal{C}})_{\vec k_2\lambda_2, \vec k\bar\lambda}
  a^\dagger_{\vec {k}_1\lambda_1}
  a_{\vec {k}_2\lambda_2}B_{\vec q}+h.c.  
\end{align}
Here the coupling matrix is assumed to be a real symmetric matrix which is momentum-independent, i.e. $g_\alpha
=\begin{pmatrix}
  g^c_\alpha & g^{cv}_\alpha \\
  g^{vc}_\alpha & g^v_\alpha 
\end{pmatrix}$ 
with $g^{cv}=g^{vc}$, and $[g_\alpha,g_{\alpha'}]=0$; the matrix element of $\mathcal C$ is $\mathcal C_\vec{k\lambda,k'\lambda'}\equiv \sum_\alpha g^{\lambda\lambda'}_{\alpha} Q_{\vec{k-k'},\alpha}$.

\subsection{Green's function}\label{ap_gf2}

In the following, we calculate the absorption coefficient through the greater part of the retarded four-point correlation function. We parallel the previous derivation and resort to the Polaron picture by performing a unitary transformation so that the electron-phonon coupling is incorporated nonperturbatively. From now on, $H$ denotes the full Hamiltonian without $H_{e-em}$ and $H^0_{em}$, i.e. $H=H^0_e+H^0_{ph}+H_{e-e}+H_{e-ph}$, unless otherwise stated. Using Eq.(\ref{eq:a}) and Eq.(\ref{eq:a_dagger}), gives
\begin{align}
  (e^{\textup{ad}_S}a^{\dagger}_{\vec k \lambda})(t)
  &=\sum_{\vec p}e^{i(\bar H_{pol}+\bar H_{pol-pol})t}
    a^{\dagger}_{\vec k \lambda}e^{-i(\bar H_{pol}+\bar H_{pol-pol})t} 
    e^{i\bar H^0_{ph}t}(e^{\mathcal C^\lambda})_{\vec{kp}}e^{-i\bar H^0_{ph}t}\nonumber 
\end{align}
from which we obtain the transformed greater four-point correlation function 
\begin{align}
  P^R_>(t)
  &=-i\frac{\Theta(t)}{N}\sum_{\vec {k,k'},\{\vec k_i\lambda_i\}}
  \left\langle
  a^\dagger_{\vec k_1 \lambda_1}(t) 
  a_{\vec{k}_2\lambda_2}(t) 
  a^\dagger_{\vec k_3 \lambda_3}(0) 
  a_{\vec k_4 \lambda_4} (0)
  \right\rangle\\
  &\qquad\times
  \left\langle 
  (e^{\mathcal{C}(t)})_{\vec{k}v,\vec{k}_1\lambda_1} 
  (e^{-\mathcal{C}(t)})_{\vec{k}_2\lambda_2,\vec{k}c} 
  (e^{\mathcal{C}(0)})_{\vec{k}'c,\vec{k}_3\lambda_3} 
  (e^{-\mathcal{C}(0)})_{\vec{k}_4\lambda_4,\vec{k}'v} 
  \right\rangle
\end{align}
where polaron operator evolves according to $a^{(\dagger)}_{\vec k \lambda}(t)=e^{i\bar H_{p}t} a^{(\dagger)}_{\vec k \lambda}e^{-i\bar H_{p}t}$ and $\bar H_p=\bar H_{pol}+\bar H_{pol-pol}$, and similarly $\mathcal C(t)=e^{i\bar H_{ph}t} \mathcal C(0)e^{-i\bar H_{ph}t}$ with matrix elements being $\mathcal C(t)_{\vec{k}\lambda\vec{k'}\lambda'}=\sum_\alpha g^{\lambda\lambda'}_{\alpha}(e^{i\Omega_\alpha t}D^\dagger_{\vec{k-k'},\alpha}-e^{-i\Omega_\alpha t}D_{\vec{k'-k},\alpha})$. In polaron picture, the polaronic and phononic operators separate as before.

However, at finite temperatures, the previous method of Feynman disentangling of operators is not applicable due to the mixing of the band indices and the momentum indices. Therefore, we need to first disentangle these indices. For simplicity, we only consider one phonon brach so that $\alpha$ index can be removed. The Fröhlich coupling matrix for antisymmetric intraband and symmetric interband coupling takes the form $g=g_1\sigma_1+g_3\sigma_3$, where $\sigma_i$ are Pauli matrices. We also define the norm $|g|=\sqrt{g_1^2+g_3^2}$ and $\hat g=g/|g|$.

We now consider the matrix $e^{\mathcal{C}}$, where $\mathcal{C}=g\otimes Q$
\begin{align}
  e^{\mathcal{C}}=\sum_{n=0}^\infty
  \left[
    \frac{1}{(2n)!}g^{2n}\otimes Q^{2n}
    +
    \frac{1}{(2n+1)!}g^{2n+1}\otimes Q^{2n+1}
  \right]
\end{align}
Notice that 
\begin{align}
  g^{2n}= |g|^{2n}\mathbb{I},\qquad
  g^{2n+1}= |g|^{2n+1}\hat g,
\end{align}
where $\mathbb I$ is the identity matrix in the space of band indices.
Defining $\mathcal C_g = |g|Q$ yielding
\begin{align}
  e^{\mathcal{C}}
  &=\mathbb{I}\otimes \cosh(\mathcal C_g)
  +\hat g\otimes \sinh(\mathcal C_g)\nonumber\\
  &=\frac{\mathbb{I}+\hat g}{2}e^{\mathcal C_g}
  +\frac{\mathbb{I}-\hat g}{2}e^{-\mathcal C_g}
  =\sum_{\tau=\pm}\frac{\mathbb{I}+\tau\hat g}{2}e^{\tau \mathcal C_g}.
\end{align}
Therefore, 
\begin{align}
  \sum_{\vec{kk'}}
  &\left\langle 
  (e^{\mathcal{C}(t)})_{\vec{k}v,\vec{k}_1\lambda_1} 
  (e^{-\mathcal{C}(t)})_{\vec{k}_2\lambda_2,\vec{k}c} 
  (e^{\mathcal{C}(0)})_{\vec{k}'c,\vec{k}_3\lambda_3} 
  (e^{-\mathcal{C}(0)})_{\vec{k}_4\lambda_4,\vec{k}'v} 
  \right\rangle\nonumber\\
  &=\sum_{\{\tau_i=\pm\}}
  \left(\frac{\mathbb{I}+\tau_1\hat g}{2}\right)_{v\lambda_1}
  \left(\frac{\mathbb{I}+\tau_2\hat g}{2}\right)_{\lambda_2c}
  \left(\frac{\mathbb{I}+\tau_3\hat g}{2}\right)_{c\lambda_3}
  \left(\frac{\mathbb{I}+\tau_4\hat g}{2}\right)_{\lambda_4v}
  \nonumber\\
  &\qquad
  \left\langle 
  (e^{(\tau_1-\tau_2)\mathcal{C}_g(t)})_{\vec{k}_2,\vec{k}_1} 
  (e^{(\tau_3-\tau_4)\mathcal{C}_g(0)})_{\vec{k}_4,\vec{k}_3} 
  \right\rangle.
\end{align}
The Feynman disentangling method is now applicable to the above expression. The thermal average of phonon operators is
\begin{align}
  \left\langle 
  (e^{-\tau_{21}\mathcal{C}_g(t)})_{\vec{k}_2,\vec{k}_1} 
  (e^{\tau_{34}\mathcal{C}_g(0)})_{\vec{k}_4,\vec{k}_3} 
  \right\rangle
  =e^{-\frac{\tau_{21}^2+\tau_{34}^2}{2}\tilde{G}_g(T)}
  (e^{\tau_{21}\tau_{34} G_g(T,t)})_{\vec k_1 \vec k_2}\delta_{\vec k_1-\vec k_2,\vec k_4-\vec k_3},
\end{align}
where $\tau_{ij}=\tau_i-\tau_j$ and $\tilde{G}_g(T)=N[2n_{B}(\Omega,T)+1]|g|^2$, $N$ is the number of unit cells and the $N\times N$ matrix $G_g(T,t)$ is defined through
\begin{align}
  [G_g(T,t)]_{\vec k\vec k'}=  |g|^2[n_{B}(\Omega,T)e^{i\Omega t}
  +(n_{B}(\Omega,T)+1)e^{-i\Omega t}].
\end{align}
Defining the matrix elements of $\mathcal M^{\tau_1\tau_2}_{vc} (\hat g)$ as 
$\mathcal M^{\tau_1\tau_2}_{vc;\lambda_1\lambda_2} (\hat g)=
\left(\frac{\mathbb{I}+\tau_1\hat g}{2}\right)_{v\lambda_1}
\left(\frac{\mathbb{I}+\tau_2\hat g}{2}\right)_{\lambda_2c}$ and 
$\mathcal M^{\tau_3\tau_4}_{cv;\lambda_3\lambda_4} (\hat g)=
\left(\frac{\mathbb{I}+\tau_3\hat g}{2}\right)_{c\lambda_3}
  \left(\frac{\mathbb{I}+\tau_4\hat g}{2}\right)_{\lambda_4v}$,
we have
\begin{align}
  &\sum_{\vec{kk'}}
  \left\langle 
  (e^{\mathcal{C}(t)})_{\vec{k}v,\vec{k}_1\lambda_1} 
  (e^{-\mathcal{C}(t)})_{\vec{k}_2\lambda_2,\vec{k}c} 
  (e^{\mathcal{C}(0)})_{\vec{k}'c,\vec{k}_3\lambda_3} 
  (e^{-\mathcal{C}(0)})_{\vec{k}_4\lambda_4,\vec{k}'v} 
  \right\rangle\nonumber\\
  &=\sum_{\{\tau_i=\pm\}}
  \mathcal M^{\tau_3\tau_4}_{cv;\lambda_3\lambda_4} (\hat g)
  \mathcal M^{\tau_1\tau_2}_{vc;\lambda_1\lambda_2} (\hat g)
  e^{-\frac{\tau_{21}^2+\tau_{34}^2}{2}\tilde{G}_g(T)}
  (e^{\tau_{21}\tau_{34} G_g(T,t)})_{\vec k_1,\vec k_2}
  \delta_{\vec k_1-\vec k_2,\vec k_4-\vec k_3}.
\end{align}
To further simply, we define $\cos\theta=g_3/|g|$, $\sin\theta=g_1/|g|$ and assign values $c\equiv+1$, $v\equiv-1$, thus $\lambda=\pm1$, leading to
\begin{align}
  \left(\frac{\mathbb{I}+\tau\hat g}{2}\right)_{\lambda\lambda'}
  &=\sgn(\tau^+ +\lambda\lambda')\cos^{1+\frac{\tau}{2}(\lambda+\lambda')}\frac{\theta}{2}
  \sin^{1-\frac{\tau}{2}(\lambda+\lambda')}\frac{\theta}{2}
  \\
  \mathcal M^{\tau_1\tau_2}_{vc;\lambda_1\lambda_2} (\hat g)
  &=\sgn[(\tau^+_1-\lambda_1)(\tau^+_2+\lambda_2)]\cos^{X^{12}_+}\frac{\theta}{2}
  \sin^{X^{12}_-}\frac{\theta}{2}
\end{align}
where $\tau^+=\tau+0^+$ and $\sgn(\tau^+ +\lambda\lambda')=-1$ only when $\tau=-1,\lambda\lambda'=-1$ and equals one for all other cases and $X^{12}_{\pm}=2\pm\frac{\tau_{21}}{2}\pm\frac{1}{2}\sum_{i=1}^2\lambda_i\tau_i$.

For the thermal average of four-polaron operators 
$ \left\langle
a^\dagger_{\vec k_1 \lambda_1}(t) 
a_{\vec{k}_2\lambda_2}(t) 
a^\dagger_{\vec k_3 \lambda_3}(0) 
a_{\vec k_4 \lambda_4} (0)
\right\rangle$, 
we first note that 
\begin{align}
  a^\dagger_{\vec k_1 \lambda_1}(t) 
  a_{\vec{k}_2\lambda_2}(t) 
  =e^{it\cdot\textup{ad}_{\bar H_p}}
  a^\dagger_{\vec k_1 \lambda_1}
  a_{\vec{k}_2\lambda_2}.
\end{align}
For a single phonon branch with coupling $g=g_1\sigma_1+g_3\sigma_3$, 
\begin{align}
  H_p=\sum_{\vec{p}\lambda}e_{\vec p\lambda}a^\dagger_{\vec p\lambda}a_{\vec p\lambda}+\frac{1}{2}\sum_{\vec{kk'q}\{\lambda_i \} } 
  V^{\lambda_1\lambda_2  \lambda_4\lambda_3}_{\vec q}
  a^\dagger_{\vec{k-q}\lambda_1}
  a^\dagger_{\vec{k'+q}\lambda_2}
  a_{\vec k'\lambda_3}
  a_{\vec k\lambda_4}
\end{align}
where $V^{\lambda_1\lambda_2  \lambda_4\lambda_3}_{\vec q}=V_\vec{q} \delta_{\lambda_2,\lambda_3}\delta_{\lambda_1,\lambda_4}-2\Omega g^{\lambda_4\lambda_1}g^{\lambda_2\lambda_3}$ with the symmetry that 
$V^{\lambda_1\lambda_2  \lambda_4\lambda_3}_{\vec q}
=V^{\lambda_2\lambda_1  \lambda_3\lambda_4}_{\vec q}
=V^{\lambda_4\lambda_3\lambda_1\lambda_2}_{\vec q}$ and $V_{\vec{q}}=V_{-\vec{q}}$.
The adjoint action of $\bar H_p$ on the polaron bilinear operators in general leads to hierarchy problems in which truncation should be taken at certain order. We instead  perform mean-field approximation, which yields
\begin{align}
  \textup{ad}_{\bar H_p} 
  a^\dagger_{\vec k_1 \lambda_1}
  a_{\vec{k}_2\lambda_2}
  &\approx
   -\sum_{\vec k'_1\lambda_1'\vec k'_2\lambda_2'} 
   K_{\vec k_1 \lambda_1 \vec k_2 \lambda_2;\vec k'_1 \lambda_1'\vec k'_2 \lambda_2'}
  a^\dagger_{\vec k'_1 \lambda_1'}a_{\vec k'_2 \lambda_2'}
  \\
   K_{\vec k_1 \lambda_1 \vec k_2 \lambda_2;\vec k'_1 \lambda_1'\vec k'_2 \lambda_2'}
  &=\left[
    \tilde e_{\vec k_2\lambda_2\lambda_2'}
    \delta_{\lambda_1 \lambda_1'}
    -
    \tilde e_{\vec k_1\lambda_1\lambda_1'}
   \delta_{\lambda_2 \lambda_2'}
  \right]\delta_{\vec k_1 \vec k_1'}\delta_{\vec k_2 \vec k_2'}
  +\mathcal K_{\vec k_1 \lambda_1 \vec k_2 \lambda_2;\vec k'_1 \lambda_1'\vec k'_2 \lambda_2'}
  \\
  \mathcal K_{\vec k_1 \lambda_1 \vec k_2 \lambda_2;\vec k'_1 \lambda_1'\vec k'_2 \lambda_2'}
  &=(n_{\vec k_2\lambda_2}-n_{\vec k_1\lambda_1})
  \left(
    V_{\vec k_1-\vec k_1'}^{\lambda_1'\lambda_2\lambda_1\lambda_2'} 
    -
    V_{\vec k_1-\vec k_2}^{\lambda_2\lambda_1'\lambda_1\lambda_2'}
  \right)
  \delta_{\vec k_1-\vec k'_1,\vec k_2-\vec k'_2},
\end{align}
where $\tilde e_{\vec k\lambda_1\lambda_1'}=e_{\vec k_1\lambda_1} \delta_{\lambda_1 \lambda_1'}+\Sigma_{\vec k_1\lambda_1\lambda_1'}$ and $\Sigma_{\vec k_1\lambda_1\lambda_1'}=\sum_{\vec{q}\lambda}(V_{\vec{0}}^{\lambda\lambda_1'\lambda\lambda_1}-V_{\vec{q}}^{\lambda\lambda_1'\lambda_1\lambda})n_{\vec{k}_1-\vec{q}\lambda}$ is the Hartree-Fock energy shift. For materials with band gaps larger than 1eV, we can further make the ``large-band-gap approximation" for the fermionic occupation number, i.e. $n_{\vec kc}\approx 0\ \forall c, n_{\vec kv}\approx 1\ \forall v$, which remains valid for temperatures up to $10^4$K for materials without doping. Using this approximation, the self energy is diagonal  
$\Sigma_{\vec k_1\lambda_1\lambda_1'
}=\sum_{\vec{q}}(V_\vec{0}^{v\lambda_1'v\lambda_1}
-V_{\vec{q}}^{v\lambda_1'\lambda_1v})=\text{diag}(\Sigma_{\vec k_1 c},\Sigma_{\vec k_1 v})$ with $\Sigma_{\vec k_1 c}=N(V_\vec{0}+2|g|^2\Omega)$ and $\Sigma_{\vec k_1 v}=NV_\vec{0}-\sum_{\vec{q}}V_\vec{q}$. Moreover, we can rewrite the kernel as
\begin{align}
   K_{\vec k_1 \lambda_1 \vec k_2 \lambda_2;\vec k'_1 \lambda_1'\vec k'_2 \lambda_2'}
  &=\sum_{i=0}^2\mathcal K^{(i)}_{\vec k_1 \lambda_1 \vec k_2 \lambda_2;\vec k'_1 \lambda_1'\vec k'_2 \lambda_2'}
  \\
  \mathcal K^{(0)}_{\vec k_1 \lambda_1 \vec k_2 \lambda_2;\vec k'_1 \lambda_1'\vec k'_2 \lambda_2'}
  &=(\tilde e_{\vec k_2\lambda_2}-
  \tilde e_{\vec k_1\lambda_1})
  \delta_{\lambda_1 \lambda_1'}
  \delta_{\lambda_2 \lambda_2'}
  \delta_{\vec k_1 \vec k_1'}\delta_{\vec k_2 \vec k_2'}
  \\
  \mathcal K^{(1)}_{\vec k_1 \lambda_1 \vec k_2 \lambda_2;\vec k'_1 \lambda_1'\vec k'_2 \lambda_2'}
  &=-\delta_{\lambda_2c}\delta_{\lambda_1v}
  \left(
    V_{\vec k_1-\vec k_1'}^{\lambda_1'cv\lambda_2'} 
    -
    V_{\vec k_1-\vec k_2}^{c\lambda_1'v\lambda_2'}
  \right)
  \delta_{\vec k_1-\vec k'_1,\vec k_2-\vec k'_2}
  \nonumber
  \\
  &=-\delta_{\lambda_2c}\delta_{\lambda_1v}
  \delta_{\lambda_1'v}\delta_{\lambda_2'c}
  (V_{\vec k_1-\vec k_1'}+2|g|^2\Omega)
  \delta_{\vec k_1-\vec k'_1,\vec k_2-\vec k'_2}
  \\
  \mathcal K^{(2)}_{\vec k_1 \lambda_1 \vec k_2 \lambda_2;\vec k'_1 \lambda_1'\vec k'_2 \lambda_2'}
  &= \delta_{\lambda_2v}\delta_{\lambda_1c}
  \left(
    V_{\vec k_1-\vec k_1'}^{\lambda_1'vc\lambda_2'} 
    -
    V_{\vec k_1-\vec k_2}^{v\lambda_1'c\lambda_2'}
  \right)\nonumber\\
  &= \delta_{\lambda_2v}\delta_{\lambda_1c}
  \delta_{\lambda_2'v}\delta_{\lambda_1'c}
  (V_{\vec k_1-\vec k_1'}+2|g|^2\Omega)
  \delta_{\vec k_1-\vec k'_1,\vec k_2-\vec k'_2}.
\end{align}
Importantly, $\mathcal{K}^{(i)}$ preserves the band indices during each scattering process, i.e. $\lambda_1'=\lambda_1$ and $\lambda_2'=\lambda_2$. This property significantly simplifies the subsequent calculations by restricting the scattering processes to inter-band transitions, as the intra-band scattering potential is suppressed by the $n_{\vec k_2\lambda}-n_{\vec k_1\lambda}$ factor in the case without doping. The time evolution is 
\begin{align}
  a^\dagger_{\vec k_1 \lambda_1}(t) 
  a_{\vec{k}_2\lambda_2}(t) 
  =\sum_{\vec k'_1\lambda_1'\vec k'_2\lambda_2'} 
  (e^{-itK})_{\vec k_1 \lambda_1 \vec k_2 \lambda_2;\vec k'_1 \lambda_1'\vec k'_2 \lambda_2'}
  a^\dagger_{\vec k'_1 \lambda_1'}a_{\vec k'_2 \lambda_2'}.
\end{align}
The time evolution of the thermal average of four-polaron operators follows from 
\begin{align}
\langle a^\dagger_{\vec k'_1 \lambda_1'}a_{\vec k'_2 \lambda_2'}
a^\dagger_{\vec k_3 \lambda_3}a_{\vec k_4 \lambda_4} \rangle
&=n_{\vec{k}_4\lambda_4}(1-n_{\vec{k}_3\lambda_3})
\delta_{\vec k_2'\vec k_3}\delta_{\vec k_1'\vec k_4}
\delta_{\lambda_2'\lambda_3}\delta_{\lambda_1'\lambda_4}
+n_{\vec k_3\lambda_3}n_{\vec k_1'\lambda_1'}
\delta_{\vec k_3\vec k_4}\delta_{\vec k_1'\vec k_2'}
\delta_{\lambda_1'\lambda_2'}\delta_{\lambda_3\lambda_4}\nonumber\\
&\approx
\delta_{\lambda_4v}\delta_{\lambda_3c}
\delta_{\vec k_2'\vec k_3}\delta_{\vec k_1'\vec k_4}
\delta_{\lambda_2'\lambda_3}\delta_{\lambda_1'\lambda_4}
+
\delta_{\lambda_3v}\delta_{\lambda_1'v}
\delta_{\vec k_3\vec k_4}\delta_{\vec k_1'\vec k_2'}
\delta_{\lambda_1'\lambda_2'}\delta_{\lambda_3\lambda_4}
\end{align}
which leads to 
\begin{align}
  &\left\langle
  a^\dagger_{\vec k_1 \lambda_1}(t) 
  a_{\vec{k}_2\lambda_2}(t) 
  a^\dagger_{\vec k_3 \lambda_3}(0) 
  a_{\vec k_4 \lambda_4} (0)
  \right\rangle
  \nonumber\\
  &=
  (e^{-itK})_{\vec k_1 \lambda_1 \vec k_2 \lambda_2;\vec k_4 \lambda_4\vec k_3 \lambda_3}
  \delta_{\lambda_4v}\delta_{\lambda_3c}\nonumber\\
  &\qquad
  +\sum_{\vec k'_1} 
  (e^{-itK})_{\vec k_1 \lambda_1 \vec k_2 \lambda_2;\vec k'_1 v\vec k'_1 v}
  \delta_{\lambda_3v}
  \delta_{\lambda_4v}
  \delta_{\vec k_3\vec k_4}
  \nonumber\\
  &=
  (e^{-it(\mathcal K^{(0)}+\mathcal K^{(1)})})_{\vec k_1 \lambda_1 \vec k_2 \lambda_2;\vec k_4 \lambda_4\vec k_3 \lambda_3}
  \delta_{\lambda_4v}\delta_{\lambda_3c}
  \delta_{\lambda_1v}\delta_{\lambda_2c}\nonumber\\
  &\qquad
  +\sum_{\vec k'_1} 
  (e^{-it\mathcal K^{(0)}})_{\vec k_1 \lambda_1 \vec k_2 \lambda_2;\vec k'_1 v\vec k'_1 v}
  \delta_{\lambda_1v}
  \delta_{\lambda_2v}
  \delta_{\lambda_3v}
  \delta_{\lambda_4v}
  \delta_{\vec k_1\vec k_2}
  \delta_{\vec k_3\vec k_4}.
\end{align}
The second term above vanishes as $\mathcal K^{(0)}_{\vec k_1v\vec k_1v;\vec k_1'v\vec k_1'v}=0$. For later convenience we further define the band-independent part as
\begin{align}
  \mathcal K_{\vec k_1\vec k_2;\vec k_1'\vec k_2'}\equiv(\mathcal K^{(0)}+\mathcal K^{(1)})_{\vec k_1 v \vec k_2 c;\vec k'_1 v\vec k'_2 c}
  =(\tilde e_{\vec k_2c}-
  \tilde e_{\vec k_1v})
  \delta_{\vec k_1 \vec k_1'}\delta_{\vec k_2 \vec k_2'}
  -
  (V_{\vec k_1-\vec k_1'}+2|g|^2\Omega)
  \delta_{\vec k_1-\vec k'_1,\vec k_2-\vec k'_2}.
\end{align}
\emph{Wannier equation.} We identify the eigenvalue equation of the kernel $K$ as the Wannier equation, i.e.
\begin{align}
  \mathcal K|\psi_{\mu}\rangle=E_\mu|\psi_{\mu}\rangle,
\end{align}
where $\mu$ is a collective index including continuous center-of-mass momentum indices labeled by $\vec Q$ and a set of discrete quantum numbers $\nu$, i.e. $\mu=(\vec Q,\nu)$. The eigenstate is a tensor product of c.o.m wavefunction and relative-motion wave function $|\psi_{\vec Q\nu}\rangle=|\vec Q\rangle\otimes|\psi_\nu\rangle$.

\vskip 1cm

\subsubsection{Absorption spectrum at finite temperatures}
We finally end up with 
\begin{align}
  P^{R}_>(t)
  &=-i\frac{\Theta(t)}{N}
  \sum_{\{\vec k_i\}}
  \sum_{\{\tau_i=\pm\}}
  \mathcal M^{\tau_3\tau_4}_{cv;cv} (\hat g)
  \mathcal M^{\tau_1\tau_2}_{vc;vc} (\hat g)
  (e^{\tau_{21}\tau_{34} G_g(T,t)})_{\vec k_1,\vec k_2}
  (e^{-it\mathcal K})_{\vec k_1 \vec k_2 ;\vec k_4 \vec k_3 }
  e^{-\frac{\tau_{21}^2+\tau_{34}^2}{2}\tilde{G}_g(T)}
  \nonumber\\
  &=-i\frac{\Theta(t)}{N}
  \sum_{\{\tau_i=\pm\}}
  \mathcal M^{\tau_3\tau_4}_{cv;cv} (\hat g)
  \mathcal M^{\tau_1\tau_2}_{vc;vc} (\hat g)
  \left\langle
  e^{\tau_{21}\tau_{34} G_g(T,t)}
  \left|
  e^{-it\mathcal K}
  \right|
  e^{-\frac{\tau_{21}^2+\tau_{34}^2}{2}\tilde{G}_g(T)}
  \right\rangle.
\end{align}
Here we explain the notation: the inner product of two bilocal functions in momentum space $f$, $g$  is given by $\langle f|g\rangle=\sum_{k_1k_2}f_{k_1,k_2}g_{k_1,k_2}$, and the action of a linear functional $A$ is $A|f\rangle_{k_1,k_2}=\sum_{k_3,k_4}A_{k_1k_2;k_3k_4}f_{k_3,k_4}$, and $|e^{-\frac{\tau_{21}^2+\tau_{34}^2}{2}\tilde{G}_g(T)}\rangle$ is a constant function with value $e^{-\frac{\tau_{21}^2+\tau_{34}^2}{2}\tilde{G}_g(T)}$ in momentum space. The correspondence between the values of $\tau_{21}$ and $\tau_{34}$ with those of $\tau_{1}$,$\tau_{2}$,$\tau_{3}$, and $\tau_{1}$ are shown in Table. \ref{tab:tau_correspondence}.

\begin{table}
  \centering
  \caption{Correspondence between $(\tau_{21}, \tau_{34})$ and $(\tau_1, \tau_2, \tau_3, \tau_4)$.}
  \label{tab:tau_correspondence}
  \begin{tabular}{cc}
  \toprule
  $(\tau_{21},\, \tau_{34})$ & $(\tau_1,\, \tau_2,\, \tau_3,\, \tau_4)$ \\
  \midrule
  $(0,\phantom{-} 0)$  & $(+1,+1,+1,+1),\ (+1,+1,-1,-1),\ (-1,-1,+1,+1),\ (-1,-1,-1,-1)$ \\[4pt]
  $(0,\phantom{-} 2)$  & $(+1,+1,+1,-1),\ (-1,-1,+1,-1)$ \\[4pt]
  $(0,          -2)$   & $(+1,+1,-1,+1),\ (-1,-1,-1,+1)$ \\[4pt]
  $(-2,\phantom{-}0)$  & $(+1,-1,+1,+1),\ (+1,-1,-1,-1)$ \\[4pt]
  $(-2,\phantom{-}2)$  & $(+1,-1,+1,-1)$ \\[4pt]
  $(-2,         -2)$   & $(+1,-1,-1,+1)$ \\[4pt]
  $(\phantom{-}2,\phantom{-}0)$ & $(-1,+1,+1,+1),\ (-1,+1,-1,-1)$ \\[4pt]
  $(\phantom{-}2,\phantom{-}2)$ & $(-1,+1,+1,-1)$ \\[4pt]
  $(\phantom{-}2,         -2)$  & $(-1,+1,-1,+1)$ \\
  \bottomrule
  \end{tabular}
\end{table}

The effective coupling matrix is
\begin{align}
  \mathcal M^{\tau_1\tau_2}_{vc;vc} (\hat g)
  =
  \cos^{(2+\tau_{21})}\frac{\theta}{2}
  \sin^{(2-\tau_{21})}\frac{\theta}{2}
  \qquad
  \mathcal M^{\tau_3\tau_4}_{cv;cv} (\hat g)
  =
  \cos^{(2+\tau_{34})}\frac{\theta}{2}
  \sin^{(2-\tau_{34})}\frac{\theta}{2}
\end{align}
where $\theta=\cos^{-1}(g_3/|g|)$. We then have
\begin{align}\label{eq:P_model2}
  \frac{P^{R}_>(t)\big|_{t>0}}{-i/N}
  &=
  \sum_{\{\tau_i=\pm\}}
  \cos^{(4+\tau_{21}+\tau_{34})}\frac{\theta}{2}
  \sin^{(4-\tau_{21}-\tau_{34})}\frac{\theta}{2}
  \left\langle
  e^{\tau_{21}\tau_{34} G_g(T,t)}
  \left|
  e^{-it\mathcal K}
  \right|
  e^{-\frac{\tau_{21}^2+\tau_{34}^2}{2}\tilde{G}_g(T)}
  \right\rangle
  \nonumber\\
  &=
  4
  \cos^{(4)}\frac{\theta}{2}
  \sin^{(4)}\frac{\theta}{2}
  \left\langle
  \mathcal I
  \left|
  e^{-it\mathcal K}
  \right|
  1
  \right\rangle
  +
  2
  \cos^{(4)}\frac{\theta}{2}
  \sin^{(4)}\frac{\theta}{2}
  \left\langle
  e^{-4 G_g(T,t)}
  \left|
  e^{-it\mathcal K}
  \right|
  e^{-4\tilde{G}_g(T)}
  \right\rangle
  \nonumber\\
  &\qquad+
  4
  \left(
    \cos^{(6)}\frac{\theta}{2}
    \sin^{(2)}\frac{\theta}{2}
    +
    \cos^{(2)}\frac{\theta}{2}
    \sin^{(6)}\frac{\theta}{2}
  \right)
  \left\langle
  \mathcal I
  \left|
  e^{-it\mathcal K}
  \right|
  e^{-2\tilde{G}_g(T)}
  \right\rangle
  \nonumber\\
  &\qquad+
  \left(\sin^{(8)}\frac{\theta}{2}+\cos^{(8)}\frac{\theta}{2}\right)
  \left\langle
  e^{4 G_g(T,t)}
  \left|
  e^{-it\mathcal K}
  \right|
  e^{-4\tilde{G}_g(T)}
  \right\rangle.
\end{align}
Here $\ket{1}$ is constant function in the momentum space with value 1 and $\mathcal I$ is the identity matrix in the momentum space.
When the coupling matrix is diagonal i.e. $g=g_3\sigma_3$, $\theta=0$ and only the term $\cos^{(8)}\frac{\theta}{2}\left\langle e^{4 G_g(T,t)}\left|e^{-it\mathcal K}\right|e^{-4\tilde{G}_g(T)}\right\rangle$ is non-zero, which reduces to the previous result. The factor of 4 is due to the definition $g_3=g^{cv}/2$, where $g^{cv}=g^c-g^v$ is the effective coupling in Model I. We also rescale the coupling matrix by $g\rightarrow g/\sqrt{N}$ for comparison with the previous result.

We first consider the case where there is only one exciton species.
Assuming the excitonic dispersion is relatively flat $E_{\vec{Q}\nu}\approx E_{\vec{0}\nu}$, we have
\begin{align}
  \left\langle\mathcal I\left|e^{-it\mathcal K}\right|1\right\rangle
  &=N^2_\Omega \sum_{\nu} e^{-iE_{\vec 0\nu}t}|\psi_{\nu}(\vec 0)|^2
  \\
  \left\langle
  \mathcal I
  \left|
  e^{-it\mathcal K}
  \right|
  e^{-2\tilde{G}_g(T)}
  \right\rangle
  &= N^2_\Omega \sum_{\nu}e^{-iE_{\vec 0\nu}t}|\psi_{\nu}(\vec 0)|^2 e^{-2\tilde{G}_g(T)}
\end{align}
\begin{align}
  &\left\langle
  e^{4 G_g(T,t)}
  \left|
  e^{-it\mathcal K}
  \right|
  e^{-4\tilde{G}_g(T)}
  \right\rangle
  \nonumber\\
  =&N\sum_{\vec Q\nu} |\psi_\nu(\vec 0)|^2
  e^{4|g|^2\left[n_B(\Omega,T)(e^{i\Omega t}-1)
  +(n_B(\Omega,T)+1)(e^{-i\Omega t}-1)\right]-iE_{\vec Q\nu}t}\nonumber\\
  =&N\sum_{\vec Q\nu} |\psi_\nu(\vec 0)|^2
  \sum_{m\in\mathbb Z}e^{\frac{1}{2}m\beta\Omega-im\Omega t -iE_{\vec Q\nu}t}
  I_m\left(\frac{4|g|^2}{\sinh(\beta\Omega/2)}\right)e^{-4\tilde{G}_g(T)}
  \\
  &\left\langle
  e^{-4 G_g(T,t)}
  \left|
  e^{-it\mathcal K}
  \right|
  e^{-4\tilde{G}_g(T)}
  \right\rangle
  \nonumber\\
  =&N\sum_{\vec Q\nu} |\psi_\nu(\vec 0)|^2
  e^{-4|g|^2\left[n_B(\Omega,T)(e^{i\Omega t}+1)
  +(n_B(\Omega,T)+1)(e^{-i\Omega t}+1)\right]-iE_{\vec Q\nu}t}\nonumber\\
  =&N\sum_{\vec Q\nu} |\psi_\nu(\vec 0)|^2
  \sum_{m\in\mathbb Z}e^{\frac{1}{2}m\beta\Omega-im\Omega t- iE_{\vec Q\nu}t}(-1)^m
  I_m\left(\frac{4|g|^2}{\sinh(\beta\Omega/2)}\right)e^{-4\tilde{G}_g(T)}
\end{align}
Therefore, using the approximation $E_{\vec{Q}\nu}\approx E_{\vec{0}\nu}$, we have
\begin{align}
  \frac{P^{R}_>(t)\big|_{t>0}}{-i/N}
  &=
  \mathcal W_0(g)
  N^2_\Omega \sum_{\nu} e^{-iE_{\vec 0\nu}t}|\psi_{\nu}(\vec 0)|^2\nonumber\\
  &\quad
  +
  \mathcal W_1(g)
  N\sum_{\vec Q\nu} |\psi_\nu(\vec 0)|^2
  \sum_{m\in\mathbb Z}e^{\frac{1}{2}m\beta\Omega-im\Omega t -iE_{\vec 0\nu}t}(-1)^m
  I_m\left(\frac{4|g|^2}{\sinh(\beta\Omega/2)}\right)e^{-4\tilde{G}_g(T)}
  \nonumber\\
  &\quad+
  \mathcal W_2(g)
  N\sum_{\vec Q\nu} |\psi_\nu(\vec 0)|^2
  \sum_{m\in\mathbb Z}e^{\frac{1}{2}m\beta\Omega-im\Omega t -iE_{\vec 0\nu}t}
  I_m\left(\frac{4|g|^2}{\sinh(\beta\Omega/2)}\right)e^{-4\tilde{G}_g(T)}
\end{align}
where the weight functions are
\begin{align}
  \mathcal W_0(g)
  &=4
  \left[
    \cos^{(4)}\frac{\theta}{2}
    \sin^{(4)}\frac{\theta}{2}
    +
    \left(
    \cos^{(6)}\frac{\theta}{2}
    \sin^{(2)}\frac{\theta}{2}
    +
    \cos^{(2)}\frac{\theta}{2}
    \sin^{(6)}\frac{\theta}{2}
    \right)e^{-2\tilde{G}_g(T)}
  \right]
  \\
  \mathcal W_1(g)
  &
  =2
  \cos^{(4)}\frac{\theta}{2}
  \sin^{(4)}\frac{\theta}{2}
  \\
  \mathcal W_2(g)
  &=\left(\sin^{(8)}\frac{\theta}{2}+\cos^{(8)}\frac{\theta}{2}\right).
\end{align}
Therefore, the Fourier transformation $P^R_>(\omega)=\int_t P^R(t)e^{i\omega^+ t}$ is given by
\begin{align}
  P^R_>(\omega,T)
  =\sum_{i=0}^{2}P^{R(i)}_>(\omega,T)
\end{align}

\begin{align}
  P^{R(0)}_>(\omega,T)/N
  &=\mathcal W_0(g) \sum_{\nu}\frac{|\psi_{\nu}(\vec 0)|^2}{\omega^+-E_{\vec{0}\nu}}
  \\
  P^{R(1)}_>(\omega,T)/N
  &=\mathcal W_1(g)e^{-4\tilde{G}_g(T)}
  \sum_{\nu} \sum_{m=-\infty}^\infty
  \frac{(-1)^m e^{\frac{1}{2}m\beta\Omega}|\psi_\nu(\vec 0)|^2}{\omega^+-m\Omega-E_{\vec{0}\nu}}
  I_m\left(\frac{4|g|^2}{\sinh(\beta\Omega/2)}\right)
  \\
  P^{R(2)}_>(\omega,T)/N
  &=\mathcal W_2(g)e^{-4\tilde{G}_g(T)}
  \sum_{\nu} \sum_{m=-\infty}^\infty
  \frac{ e^{\frac{1}{2}m\beta\Omega}|\psi_\nu(\vec 0)|^2}{\omega^+-m\Omega-E_{\vec{0}\nu}}
  I_m\left(\frac{4|g|^2}{\sinh(\beta\Omega/2)}\right).
\end{align}
When $g_3=0$, $\theta=\cos^{-1}(g_3/|g|)=0$ and only the last term $P^{R(2)}_>(\omega,T)$ will survive, which reduces to the previous result given that $\tilde{G}_g(T)=[2n_{B}(\Omega,T)+1]|g|^2$ and $|g|=g_3=g^{cv}/2$.
The generalization to multiple excitons is straightforward. One should treat $\nu$ as a collective index which include both the exciton energy levels and exciton species. Finally, the total spectrum in the main text can be obtained by taking the imaginary part of $P^R_>(\omega,T)$.

\twocolumngrid
\bibliography{References}

@article{Feldtmann,
    author = {Feldtmann, T and Kira, M and Koch, Stephan W},
	journal = {physica status solidi (b)},
	number = {2},
	pages = {332--336},
	publisher = {Wiley Online Library},
	title = {Phonon sidebands in semiconductor luminescence},
	volume = {246},
	year = {2009}
}

@book{doi:10.1142/7184,
    author = {Haug, Hartmut and Koch, Stephan W},
    title = {Quantum Theory of the Optical and Electronic Properties of Semiconductors},
    publisher = {WORLD SCIENTIFIC},
    year = {2009},
    doi = {10.1142/7184},
    address = {},
    pages = {163-187},
    edition   = {5th},
    URL = {https://www.worldscientific.com/doi/abs/10.1142/7184},
    eprint = {https://www.worldscientific.com/doi/pdf/10.1142/7184}
}

@article{PhononData,
    author = {Barati, Fatemeh and Arp, Trevor B. and Su, Shanshan and Lake, Roger K. and Aji, Vivek and van Grondelle, Rienk and Rudner, Mark S. and Song, Justin C. W. and Gabor, Nathaniel M.},
    title = {Vibronic Exciton–Phonon States in Stack-Engineered van der Waals Heterojunction Photodiodes},
    journal = {Nano Letters},
    volume = {22},
    number = {14},
    pages = {5751-5758},
    year = {2022},
    doi = {10.1021/acs.nanolett.2c00944},
        note ={PMID: 35787025},
    URL = { https://doi.org/10.1021/acs.nanolett.2c00944},
    eprint = {https://doi.org/10.1021/acs.nanolett.2c00944}
}

@article{PhysRevLett.123.247402,
  title = {Infrared Interlayer Exciton Emission in ${\mathrm{MoS}}_{2}/{\mathrm{WSe}}_{2}$ Heterostructures},
  author = {Karni, Ouri and Barr\'e, Elyse and Lau, Sze Cheung and Gillen, Roland and Ma, Eric Yue and Kim, Bumho and Watanabe, Kenji and Taniguchi, Takashi and Maultzsch, Janina and Barmak, Katayun and Page, Ralph H. and Heinz, Tony F.},
  journal = {Phys. Rev. Lett.},
  volume = {123},
  issue = {24},
  pages = {247402},
  numpages = {7},
  year = {2019},
  month = {Dec},
  publisher = {American Physical Society},
  doi = {10.1103/PhysRevLett.123.247402},
  url = {https://link.aps.org/doi/10.1103/PhysRevLett.123.247402}
}

@article{munn1985theory,
	author = {Munn, RW and Silbey, R},
	journal = {The Journal of chemical physics},
	number = {4},
	pages = {1843--1853},
	publisher = {American Institute of Physics},
	title = {Theory of electronic transport in molecular crystals. II. Zeroth order states incorporating nonlocal linear electron--phonon coupling},
	volume = {83},
	year = {1985}}

@article{antonius2022theory,
	author = {Antonius, Gabriel and Louie, Steven G},
	journal = {Physical Review B},
	number = {8},
	pages = {085111},
	publisher = {APS},
	title = {Theory of exciton-phonon coupling},
	volume = {105},
	year = {2022}
}

@article{hannewald2005nonperturbative,
	author = {Hannewald, K and Bobbert, PA},
	journal = {Physical Review B},
	number = {11},
	pages = {113202},
	publisher = {APS},
	title = {Nonperturbative theory of exciton-phonon resonances in semiconductor absorption},
	volume = {72},
	year = {2005}
}

@book{haug2009quantum,
	author = {Haug, Hartmut and Koch, Stephan W},
	publisher = {World Scientific Publishing Company},
	title = {Quantum theory of the optical and electronic properties of semiconductors},
	year = {2009}
}

@article{segall1968phonon,
	author = {Segall, B and Mahan, GD},
	journal = {Physical Review},
	number = {3},
	pages = {935},
	publisher = {APS},
	title = {Phonon-assisted recombination of free excitons in compound semiconductors},
	volume = {171},
	year = {1968}
}

@article{marsusi2012theoretical,
  title={Theoretical model obtained in momentum space for charge transport in a system consisting of noninteracting polarons},
  author={Marsusi, F and Sabbaghzadeh, J},
  journal={Physical Review B—Condensed Matter and Materials Physics},
  volume={85},
  number={11},
  pages={115302},
  year={2012},
  publisher={APS}
}

@article{RevModPhys.74.895,
  title = {Theory of ultrafast phenomena in photoexcited semiconductors},
  author = {Rossi, Fausto and Kuhn, Tilmann},
  journal = {Rev. Mod. Phys.},
  volume = {74},
  issue = {3},
  pages = {895--950},
  numpages = {0},
  year = {2002},
  month = {Aug},
  publisher = {American Physical Society},
  doi = {10.1103/RevModPhys.74.895},
  url = {https://link.aps.org/doi/10.1103/RevModPhys.74.895}
}

@article{doi:10.1126/science.1142188,
	Author = {Hohjai Lee and Yuan-Chung Cheng and Graham R. Fleming},
	Doi = {10.1126/science.1142188},
	Eprint = {https://www.science.org/doi/pdf/10.1126/science.1142188},
	Journal = {Science},
	Number = {5830},
	Pages = {1462-1465},
	Title = {Coherence Dynamics in Photosynthesis: Protein Protection of Excitonic Coherence},
	Url = {https://www.science.org/doi/abs/10.1126/science.1142188},
	Volume = {316},
	Year = {2007}}

@article{10.1063/1.3002335,
    author = {Mohseni, Masoud and Rebentrost, Patrick and Lloyd, Seth and Aspuru-Guzik, Alán},
    title = {Environment-assisted quantum walks in photosynthetic energy transfer},
    journal = {The Journal of Chemical Physics},
    volume = {129},
    number = {17},
    pages = {174106},
    year = {2008},
    month = {11},
    issn = {0021-9606},
    doi = {10.1063/1.3002335},
    url = {https://doi.org/10.1063/1.3002335},
    eprint = {https://pubs.aip.org/aip/jcp/article-pdf/doi/10.1063/1.3002335/14698253/174106\_1\_online.pdf},
}

@article{fei_nat_comm_19,
	Author = {Ma, Fei and Romero, Elisabet and Jones, Michael R. and Novoderezhkin, Vladimir I. and van Grondelle, Rienk},
	Da = {2019/02/25},
	Doi = {10.1038/s41467-019-08751-8},
	Id = {Ma2019},
	Isbn = {2041-1723},
	Journal = {Nature Communications},
	Number = {1},
	Pages = {933},
	Title = {Both electronic and vibrational coherences are involved in primary electron transfer in bacterial reaction center},
	Ty = {JOUR},
	Url = {https://doi.org/10.1038/s41467-019-08751-8},
	Volume = {10},
	Year = {2019}}

@book{harris,
  title     = "Symmetry and Spectroscopy: An Introduction to Vibrational and Electronic Spectroscopy",
  author    = "Harris, D.C. and Bertolucci, M.D.",
  year      = 1989,
  publisher = "Dover",
  address   = "New York"
}

@article{PhysRevLett.77.4062,
  title = {Generation and Relaxation of Coherent Majority Plasmons},
  author = {Cho, G. C. and Dekorsy, T. and Bakker, H. J. and H\"ovel, R. and Kurz, H.},
  journal = {Phys. Rev. Lett.},
  volume = {77},
  issue = {19},
  pages = {4062--4065},
  numpages = {0},
  year = {1996},
  month = {Nov},
  publisher = {American Physical Society},
  doi = {10.1103/PhysRevLett.77.4062},
  url = {https://link.aps.org/doi/10.1103/PhysRevLett.77.4062}
}

@inproceedings{dekorsy,
  title ="",
  author       = {Dekorsy, T. and Kim, A.M.T. and Cho, G.C> and Kohler, K and Kurz, H},
  year         = 1996,
  booktitle    = {Ultrafast Phenomena X},
  publisher    = {Springer},
  address      = {Berlin},
  series       = {Springer Series in Chemical Physics 62},
  pages        = {382},
  editor       = {Barbara, F.P. and Fujimoto, J.G. and Knox, W.H. and Zinth, W.},
  
}

@article{rivera,
	Author = {Rivera, Pasqual and Yu, Hongyi and Seyler, Kyle L. and Wilson, Nathan P. and Yao, Wang and Xu, Xiaodong},
	Da = {2018/11/01},
	Doi = {10.1038/s41565-018-0193-0},
	Id = {Rivera2018},
	Isbn = {1748-3395},
	Journal = {Nature Nanotechnology},
	Number = {11},
	Pages = {1004--1015},
	Title = {Interlayer valley excitons in heterobilayers of transition metal dichalcogenides},
	Ty = {JOUR},
	Url = {https://doi.org/10.1038/s41565-018-0193-0},
	Volume = {13},
	Year = {2018}}

@book{toyozawa,
  title     = "Optical Processes in Solids",
  author    = "Toyozawa, Y.",
  year      = 2003,
  publisher = "Cambridge University Press",
  address   = "U.K."
}

@article{PhysRevLett.94.027402,
  title = {Effect of Exciton-Phonon Coupling in the Calculated Optical Absorption of Carbon Nanotubes},
  author = {Perebeinos, Vasili and Tersoff, J. and Avouris, Phaedon},
  journal = {Phys. Rev. Lett.},
  volume = {94},
  issue = {2},
  pages = {027402},
  numpages = {4},
  year = {2005},
  month = {Jan},
  publisher = {American Physical Society},
  doi = {10.1103/PhysRevLett.94.027402},
  url = {https://link.aps.org/doi/10.1103/PhysRevLett.94.027402}
}

@article{leturcq,
	Author = {Leturcq, Renaud and Stampfer, Christoph and Inderbitzin, Kevin and Durrer, Lukas and Hierold, Christofer and Mariani, Eros and Schultz, Maximilian G. and von Oppen, Felix and Ensslin, Klaus},
	Da = {2009/05/01},
	Doi = {10.1038/nphys1234},
	Id = {Leturcq2009},
	Isbn = {1745-2481},
	Journal = {Nature Physics},
	Number = {5},
	Pages = {327--331},
	Title = {Franck--Condon blockade in suspended carbon nanotube quantum dots},
	Ty = {JOUR},
	Url = {https://doi.org/10.1038/nphys1234},
	Volume = {5},
	Year = {2009}}

@article{PhysRev.90.297,
  title = {The Motion of Slow Electrons in a Polar Crystal},
  author = {Lee, T. D. and Low, F. E. and Pines, D.},
  journal = {Phys. Rev.},
  volume = {90},
  issue = {2},
  pages = {297--302},
  numpages = {0},
  year = {1953},
  month = {Apr},
  publisher = {American Physical Society},
  doi = {10.1103/PhysRev.90.297},
  url = {https://link.aps.org/doi/10.1103/PhysRev.90.297}
}

@article{PhysRevLett.21.1637,
  title = {Exciton-Phonon Bound State: A New Quasiparticle},
  author = {Toyozawa, Y. and Hermanson, J.},
  journal = {Phys. Rev. Lett.},
  volume = {21},
  issue = {24},
  pages = {1637--1641},
  numpages = {0},
  year = {1968},
  month = {Dec},
  publisher = {American Physical Society},
  doi = {10.1103/PhysRevLett.21.1637},
  url = {https://link.aps.org/doi/10.1103/PhysRevLett.21.1637}
}
\end{document}